\documentclass[journal]{IEEEtran}
\ifCLASSINFOpdf
\else
\fi
\usepackage{graphicx,amssymb,amstext,amsmath,epsf,mathtools,indentfirst,bm,color}
\usepackage{cite}
\usepackage[small,bf]{caption}
\usepackage{algpseudocode}
\usepackage{algorithm}
\usepackage{subfigure}
\usepackage{empheq}
\usepackage{url}

\newtheorem{myprop}{\bf{Proposition}}

\newtheorem{remark}{\bf{Remark}}

\DeclareMathOperator{\tr}{tr}

\DeclareMathOperator{\cov}{cov}
\DeclareMathOperator{\diag}{{diag}}

\DeclareMathOperator*{\minimize}{\text{minimize}\,}
\DeclareMathOperator*{\st}{\text{subject to}\,}
\DeclareMathOperator*{\maximize}{\text{maximize}\,}

\newcommand{\Def}[0]{\mathrel{\mathop:}=}

\begin{document}
%
\title{Sensor Selection for   Estimation\\ with Correlated Measurement Noise}
%
%
%

\author{Sijia~Liu,~\IEEEmembership{Student Member,~IEEE,}
	Sundeep~Prabhakar~Chepuri,~\IEEEmembership{Student Member,~IEEE,}
        Makan~Fardad,~\IEEEmembership{Member,~IEEE,}
	Engin~Ma{s}azade,~\IEEEmembership{Member,~IEEE,}
        Geert~Leus,~\IEEEmembership{Fellow,~IEEE,}
	and  Pramod~K.~Varshney,~\IEEEmembership{Fellow,~IEEE}
\thanks{Copyright (c) 2015 IEEE. Personal use of this material is permitted. However, permission to use this material for any other purposes must be obtained from the IEEE by sending a request to pubs-permissions@ieee.org.}
\thanks{S. Liu, M. Fardad and P.\,K. Varshney are with the Department
of Electrical Engineering and Computer Science, Syracuse University, Syracuse,
NY, 13244 USA e-mail: \{sliu17, makan, varshney\}@syr.edu.}
{\thanks{S.\,P. Chepuri and G. Leus are with the Faculty of Electrical Engineering, Mathematics and Computer Science, Delft University of Technology, The Netherlands. 
Email:~\{s.p.chepuri, g.j.t.leus\}@tudelft.nl.}}
\thanks{E.\,\,Masazade is with the Department of Electrical and Electronics Engineering, Yeditepe University, Istanbul, 34755, Turkey e-mail: engin.masazade@yeditepe.edu.tr.}
\thanks{The work of S. Liu and P.\,K. Varshney was supported by the U.S. Air Force Office of Scientific Research (AFOSR) under grants FA9550-10-1-0458. The work of M. Fardad was supported by the National Science Foundation (NSF) under awards EAGER  ECCS-1545270 and CNS-1329885.
The work of S.\,P. Chepuri and G. Leus is supported by NWO-STW under the VICI program (10382). 
The work of E. Masazade was supported by the Scientific and Technological Research Council of Turkey (TUBITAK) under grant 113E220.}
}

%
%

\markboth{Journal of \LaTeX\ Class Files,~Vol.~xx, No.~xx, xx~2016}
{Liu \MakeLowercase{\textit{et al.}}: xxx}
%



\maketitle

\begin{abstract}
In this paper, we consider the problem of sensor selection for parameter estimation with correlated measurement noise. 
We seek optimal sensor activations by formulating an optimization problem, in which the estimation error,  given by   the trace of the  inverse of the Bayesian Fisher information  matrix, is minimized subject to   energy constraints.  Fisher information  has been widely used as an effective sensor selection criterion. However, existing  information-based sensor selection methods  are limited to the case  of uncorrelated noise or weakly correlated noise due to the use of approximate metrics. By contrast,   here
we  derive the closed form of the Fisher information matrix with respect to   sensor selection variables that is valid for any  arbitrary noise correlation regime, and  develop both a convex relaxation approach and a greedy algorithm to find  near-optimal solutions. 
We further extend our framework of sensor selection to solve the problem of  sensor scheduling, where 
a greedy algorithm is proposed to determine non-myopic (multi-time step ahead) sensor schedules. Lastly, numerical results are  provided to  illustrate the effectiveness of our approach, and to reveal the effect of noise correlation on   estimation performance. 
\end{abstract}


%
\IEEEpeerreviewmaketitle

\section{Introduction}
%
%
%
%
\IEEEPARstart{W}{ireless} sensor networks consisting of a large number of
spatially distributed sensors  have been widely used for 
environmental monitoring, source localization, and target tracking
\cite{olirod11,zoucha04,hevicyan06}.
Among the aforementioned applications, 
sensors observe an unknown parameter or state of interest and transmit their measurements to a fusion center, which then determines the global estimate.
However, due to the constraints on the communication
bandwidth and sensor battery life, it may not be
desirable to have all the  sensors report their measurements
at all time instants. Therefore, the problem of
sensor selection/scheduling arises, which aims to strike a balance between estimation accuracy and sensor activations over  space
and/or time. The importance of sensor selection  has been discussed extensively in the context of   
   various applications, such as   target tracking \cite{ZSR02}, bit allocation  \cite{MRV2012_j},  field monitoring \cite{HJB2009,liuvemfarmasvar_14}, optimal control \cite{liufarvarmas14},  power allocation  \cite{thamit08,liukarfarvar15}, optimal experiment design \cite{CV95}, and leader selection in consensus networks \cite{linfarjov14}.

In this paper, we focus on the problem of sensor selection/scheduling for parameter estimation  similar to \cite{SS2009, cheleu15, linfarjov14, MNV2012_ciss}, but with {a key difference in that the measurement noise is correlated in the problem formulation}. 
In \cite{SS2009}, the sensor selection problem
was elegantly formulated under linear measurement models, and solved via convex optimization. 
In \cite{cheleu15}, the problem of sensor selection was generalized to nonlinear measurement models  by using  the Cram\'er-Rao bound as the sensor selection criterion.
In \cite{linfarjov14},  a particular class of sensor selection problems were transformed into   the problem of leader selection in dynamical networks. 
In \cite{MNV2012_ciss}, the problem of non-myopic scheduling that determined  sensor activations over multiple future time steps was  addressed for nonlinear filtering with quantized measurements.

In the existing literature  \cite{SS2009,cheleu15,linfarjov14, MNV2012_ciss},
the study of sensor selection/scheduling problems hinges on the assumption of \textit{uncorrelated} measurement noise, which implies that sensor  observations are  {conditionally independent}  given the underlying parameter. 
Due to conditional independence, each measurement   contributes to  Fisher information {(equivalently, inverse of the Cram\'er-Rao bound on the  error covariance matrix)}   in an {additive}
manner \cite{HK_bk}. 
Accordingly, Fisher information 
becomes a \textit{linear} function with respect to the sensor selection variables (which characterize the subset of sensors we select), and  thus the resulting selection problem can be efficiently handled via convex optimization  \cite{SS2009,cheleu15}.
However, the sensed data is often  corrupted by correlated noise due to the nature of the monitored physical environment   \cite{JP2006}.
Therefore, development of  sensor selection schemes for  \textit{correlated} measurements   is a critical task.

Recently, it has been shown in \cite{rig10_thesis,jamsimmaleu14_cf_v2, shevar14, YRB2011}  that 
the presence of correlated noise makes optimal sensor selection/scheduling problems  more challenging, since 
Fisher information is no longer a linear   function with respect to the  selection variables. In \cite{rig10_thesis,jamsimmaleu14_cf_v2,shevar14}, the problem of sensor selection with correlated noise
was formulated so as to minimize an approximate expression of   the estimation error subject to an energy constraint or to minimize the energy consumption subject to an approximate estimation constraint. 
In \cite{YRB2011}, 
a  reformulation of the multi-step Kalman filter was introduced to
schedule sensors for linear dynamical systems with correlated noise.

Different from  \cite{rig10_thesis,jamsimmaleu14_cf_v2, shevar14, YRB2011}, here we derive the closed form expression of the estimation error   with respect to sensor selection variables under correlated measurement noise, which is valid for any arbitrary noise correlation matrix.
This expression is optimized via a
convex relaxation method to   determine the optimal sensor selection scheme.
We also   propose a greedy algorithm
to solve the corresponding sensor selection problem, where we show 
 that when an inactive sensor is made active,
the increase in Fisher information yields an information gain in terms of a rank-one
matrix. 
The proposed sensor selection framework  yields a more accurate sensor selection scheme than  those presented in  \cite{rig10_thesis,jamsimmaleu14_cf_v2, shevar14}, because the schemes of  \cite{rig10_thesis,jamsimmaleu14_cf_v2, shevar14} consider an approximate formulation where the noise covariance matrix is assumed to be independent of the sensor selection variables. 
We further demonstrate that the prior  formulations for sensor selection are valid only when measurement noises are {\em weakly} correlated.  In this scenario,   maximization of the trace of  the Fisher information matrix  used in  \cite{shevar14} is equivalent to the problem of maximizing a convex quadratic function over
a bounded polyhedron. The resutling problem structure enables the use of 
 optimization methods with   reduced computational complexity.

Compared to  \cite{YRB2011}, we adopt the {recursive} Fisher information  to measure the estimation performance of sensor scheduling.
However, for non-myopic (multi-time step ahead) schedules, the Fisher information matrices at consecutive time steps are coupled with each other. Due to coupling,  expressing the Fisher information matrices in a closed form is intractable. 
Therefore, we propose a greedy algorithm to seek non-myopic sensor schedules subject to cumulative and individual energy constraints.
Numerical results show that our approach yields a better estimation performance than that of  \cite{YRB2011} for state tracking.

In a preliminary version of this paper \cite{liumasfarvar15_cf}, we  studied the problem of sensor selection using  the same  framework as in \cite{rig10_thesis,jamsimmaleu14_cf_v2, shevar14}.
Compared to \cite{liumasfarvar15_cf}, we  have the following new contributions in this paper.
\begin{itemize}
\item We propose a more general but tractable sensor selection framework  that is valid for an arbitrary noise correlation matrix, and  present a suite of efficient optimization algorithms.
\item We  reveal
 drawbacks of the existing formulations in \cite{rig10_thesis,jamsimmaleu14_cf_v2, shevar14} for sensor selection, and demonstrate  their validity in only the weak noise correlation regime.
\item
We extend the proposed sensor selection approach to address the problem of non-myopic sensor scheduling, where 
the length of the time horizon and energy constraints on individual sensors are taken into account.  
\end{itemize}

The rest of the paper is organized as follows. In Section\,\ref{sec: prob}, we formulate   the problem of sensor selection with correlated noise. 
{In Section\,\ref{sec: gen_sel}, we present a convex relaxation approach and a greedy algorithm to solve the problem of sensor selection with an arbitrary noise correlation matrix.} In Section\,\ref{sec: spe_weak},  
we present   sensor selection approach  with
weakly correlated noise. In Section\,\ref{sec: non_mop}, we 
extend our framework to solve the problem of   non-myopic sensor scheduling. 
In Section\,\ref{sec: num}, we provide numerical results   to illustrate the effectiveness of our proposed methods. In Section\,\ref{sec: conc}, we summarize our work and discuss  future research directions.

\section{Problem Formulation}
\label{sec: prob}

We wish to estimate a random vector $\mathbf x \in \mathbb R^n$ with a Gaussian prior probability
density function (PDF) $\mathcal N(\boldsymbol \mu, \boldsymbol \Sigma)$. Observations of $\mathbf x$ from $m$ sensors are corrupted by    correlated measurement noise. 
To strike a balance between estimation accuracy and sensor activations, we formulate the problem of sensor selection, where the estimation error is minimized subject to a constraint on the total number of sensor activations.   

Consider a linear  system 
\begin{align}
\mathbf y = \mathbf H \mathbf x + \mathbf v,
\label{eq: meas_para_est}
\end{align}
where $\mathbf y \in \mathbb R^m$ is the measurement vector whose 
$m$th entry corresponds to a scalar observation from the $m$th sensor, 
$\mathbf H \in \mathbb R^{m \times n}$ is the observation matrix,  and $\mathbf v \in \mathbb R^m$ is the measurement noise vector that follows a 
Gaussian distribution with zero mean and  an  \textit{invertible} covariance matrix  $\mathbf R$. {We  assume that ${\mathbf x}$ and ${\mathbf v}$ are mutually independent random variables, and
  the noise covariance matrix is positive definite and thus invertible. We note  that the  noise covariance matrix is not restricted to being diagonal, so that the measurement noise could be \textit{correlated} among the sensors.
We also note that in practice,    the first two moments of $\mathbf x$ can be learnt from a parametric covariance model, such as a power exponential model  together with a training dataset of the parameter \cite{BDS2001}.

The task of sensor selection is to determine the best subset of   sensors to activate in order to minimize the estimation error, subject to a constraint on the number of activations.
We introduce a {sensor selection vector} to represent the 
activation scheme 
\begin{align}
\mathbf w = [w_{1}, w_{2}, \ldots, w_{m}]^T , ~ ~ w_{i} \in \{ 0,1 \},
\label{eq: w_variable}
\end{align}
where $w_i$ indicates whether or not the $i$th sensor is selected. 
For example, if   the $i$th sensor reports a measurement then 
 $w_i = 1$, otherwise $w_i = 0$. In other words,
the \textit{active} sensor measurements can be compactly expressed  as 
\begin{align}
\mathbf y_w= \boldsymbol \Phi_{w} \mathbf y= \boldsymbol \Phi_{w}  \mathbf H \mathbf x + \boldsymbol \Phi_{w} \mathbf v ,
\label{eq: meas_Phiw}
\end{align}
where  $\mathbf y_w \in \mathbb R^{\| \mathbf w\|_1}$ is the vector of   measurements of selected sensors, $\|\mathbf w\|_1$ is the $\ell_1$-norm of  $ \mathbf w $ which yields the total number of sensor activations, 
$ \boldsymbol \Phi_w \in \{ 0,1\}^{\| \mathbf w\|_1 \times m}$ is a submatrix of $\diag(\mathbf w)$ after   all rows
corresponding to the unselected sensors have been removed, and $\diag(\mathbf w)$  is  a diagonal matrix whose diagonal entries are given by  $\mathbf w$.  
We note that  $ \boldsymbol \Phi_{w}$ and   $\mathbf w$ are linked as below
 \begin{align}
\boldsymbol \Phi_w \boldsymbol \Phi_w^T = \mathbf I_w  ~~\text{and}~~ 
~\boldsymbol \Phi_w^T \boldsymbol \Phi_w = \diag(\mathbf w), 
\label{eq: Phi_property}
\end{align}
where  $\mathbf I_w$ denotes an identity matrix with dimension $\|\mathbf w\|_1$.

\subsection{Minimum mean-squared  estimation error}

We employ the minimum mean square  error (MMSE) estimator to   estimate  the unknown parameter under the   Bayesian setup. It is worth mentioning that the use of the Bayesian estimation framework ensures the validity of parameter estimation for an underdetermined system, in which the number of selected sensors is less than the dimension of the parameter to be estimated, namely, $\| {\mathbf w} \|_1 <  n$.

Given  the Gaussian linear measurement model \eqref{eq: meas_para_est}, the  prior distribution of the unknown parameter $\mathbf x$ and the active sensor measurements \eqref{eq: meas_Phiw}, {the error covariance matrix of the MMSE estimate of ${\mathbf x}$} is given by \cite[Theorem\,12.1]{karbook}
\begin{align}
\mathbf P_w = \left (  \boldsymbol \Sigma^{-1} + \mathbf H^T \boldsymbol \Phi_w^T \mathbf R_w^{-1} \boldsymbol \Phi_w \mathbf H \right )^{-1}, \label{eq: Pw_MSE}
\end{align}
where the matrix $\boldsymbol \Phi_w \mathbf H$ comprises rows of $\mathbf H$ for the active sensors,
  and $\mathbf R_w$ denotes the submatrix of  $\mathbf R$ after all rows and columns corresponding to the inactive sensors have been removed, i.e.,
\begin{align}
\mathbf R_w =  \boldsymbol \Phi_{w}  \mathbf R \boldsymbol \Phi_w^T.
\label{eq: Rw_sel}
\end{align}
It is clear from \eqref{eq: Pw_MSE} that due to the presence of the prior knowledge    about $\boldsymbol \Sigma$, the MSE matrix $\mathbf P_w$ is always well defined,  even if  the matrix $ \mathbf H^T \boldsymbol \Phi_w^T \mathbf R_w^{-1} \boldsymbol \Phi_w \mathbf H $ is not invertible for an underdetermined system with  $\| \mathbf w \|_1 \leq n$.

It is known from \cite{HK_bk}  that the MSE matrix $\mathbf P_w$ is the inverse of the {Bayesian} Fisher information  matrix $\mathbf J_w$ under the linear Gaussian measurement model  {with a Gaussian prior distribution}.
We thus obtain
\begin{align}
\mathbf J_w &=  \mathbf P_w ^{-1} \nonumber \\
&=  \boldsymbol \Sigma^{-1} + \mathbf H^T \boldsymbol \Phi_w^T \mathbf R_w^{-1} \boldsymbol \Phi_w \mathbf H,\label{eq: Jw}
\end{align}
where the second term is related to the sensor selection scheme.
In this paper, for clarity of presentation, we choose to work with   $\mathbf J_w$ rather than $\mathbf P_w$.

It is clear from \eqref{eq: Rw_sel} and \eqref{eq: Jw}  that the dependence of $\mathbf J_w$ on $\mathbf w$ is through $ \boldsymbol \Phi_w$. This dependency does not lend itself to easy optimization of scalar-valued functions of $\mathbf J_w$ with respect to $\mathbf w$. In what follows, we will rewrite $\mathbf J_w$ as an explicit function of the selection vector $\mathbf w$.

\subsection{Fisher information $\mathbf J_w$ as an explicit function of $\mathbf w$}

The key idea of expressing \eqref{eq: Jw} as an explicit function of $\mathbf w$ is to replace   $ \boldsymbol \Phi_w$ with $\mathbf w$ based on their relationship given by \eqref{eq: Phi_property}.  
Consider a decomposition of  the noise covariance matrix \cite{cheleu_icassp15}
\begin{align}
\mathbf R = a \mathbf I + \mathbf S, \label{eq: R_new}
\end{align}
where a positive scalar $a$ is chosen such that the matrix $\mathbf S$ is positive definite, and $\mathbf I$ is the identity matrix. 
We remark that the decomposition  given in \eqref{eq: R_new} is readily obtained through an eigenvalue decomposition of   the positive definite matrix $\mathbf R$, and it helps us in deriving  the closed form of the Fisher information matrix with respect to $\mathbf w$.

Substituting \eqref{eq: R_new} into \eqref{eq: Rw_sel}, we obtain 
\begin{align}
\mathbf R_w &=  \boldsymbol \Phi_{w} (a \mathbf I + \mathbf S) \boldsymbol \Phi_w^T   = a \mathbf I_w +  \boldsymbol \Phi_{w}  \mathbf S \boldsymbol \Phi_w^T, \label{eq: Rw_PKV}
\end{align}
where the last equality holds due to \eqref{eq: Phi_property}.

Using \eqref{eq: Rw_PKV},
we can rewrite a part of the second term on the right hand side of    \eqref{eq: Jw} as
\begin{align}
\hspace*{-0.15in}\boldsymbol \Phi_{w}^T \mathbf R_w^{-1} \boldsymbol \Phi_{w}
  & \hspace*{-0.015in} \,\,{=} \,\, \hspace*{-0.015in} \boldsymbol \Phi_{w}^T  (a \mathbf I_w +  \boldsymbol \Phi_{w}  \mathbf S \boldsymbol \Phi_w^T)^{-1} \boldsymbol \Phi_{w}   \nonumber \\
& \hspace*{-0.015in} \overset{(1)}{=} \hspace*{-0.015in} \mathbf S^{-1} \hspace*{-0.03in} - \hspace*{-0.03in} \mathbf S^{-1} (\mathbf S^{-1} \hspace*{-0.03in} + \hspace*{-0.03in} {a^{-1}\boldsymbol \Phi_w^T \boldsymbol \Phi_w})^{-1} \mathbf S^{-1} \hspace*{-0.2in} \nonumber \\
& \hspace*{-0.015in} \overset{(2)}{=} \hspace*{-0.015in} \mathbf S^{-1} \hspace*{-0.03in} - \hspace*{-0.03in} \mathbf S^{-1} (\mathbf S^{-1} \hspace*{-0.03in} + \hspace*{-0.03in} {{a}^{-1} \diag(\mathbf w)}) ^{-1} \mathbf S^{-1}, \label{eq: PhiSPhi}
\end{align}
where 
step (1) is obtained from the matrix inversion lemma\footnote{For appropriate matrices $\mathbf A$, $\mathbf B$, $\mathbf C$ and $\mathbf D$, the matrix inversion lemma states that
$(\mathbf A + \mathbf B \mathbf C \mathbf D)^{-1} = \mathbf A^{-1} - \mathbf A^{-1} \mathbf B (\mathbf C^{-1} + \mathbf D \mathbf A^{-1} \mathbf B)^{-1} \mathbf D \mathbf A^{-1}$, which yields 
$\mathbf B (\mathbf C^{-1} + \mathbf D \mathbf A^{-1} \mathbf B)^{-1} \mathbf D = \mathbf A - \mathbf A (\mathbf A + \mathbf B \mathbf C \mathbf D)^{-1} \mathbf A
$.
}, and step (2) holds due to \eqref{eq: Phi_property}.

Substituting \eqref{eq: PhiSPhi} into \eqref{eq: Jw}, the Fisher information   matrix can be expressed as
 \begin{align}
\mathbf J_w 
= &\boldsymbol \Sigma^{-1} 
+  \mathbf H^T \mathbf S^{-1}  \mathbf H \nonumber \\
&- \mathbf H^T \mathbf S^{-1} (\mathbf S^{-1} + a^{-1} \diag(\mathbf w)) ^{-1} \mathbf S^{-1} \mathbf H.
\label{eq: Jw_hat_simple}
\end{align}
It is clear from \eqref{eq: Jw_hat_simple} that the decomposition of $\mathbf R$  in \eqref{eq: R_new}, together with equations \eqref{eq: Rw_PKV}-\eqref{eq: PhiSPhi}, allows us to make explicit and isolate  the dependence of $\mathbf J_w$ on $\mathbf w$.
We also remark that
 the positive scalar $a$ and positive definite matrix $\mathbf S$ can be arbitrarily chosen, and have no effect on the performance of the sensor selection algorithms that will be proposed later on.

\subsection{Formulation of the optimal sensor selection problem}

We now state the main optimization problem considered in this work as
\begin{align}
\begin{array}{ll}
\displaystyle \minimize_{\mathbf w} 
& ~\tr(\mathbf J_w^{-1})\\
\st &~ \mathbf 1^T \mathbf w \leq s, \\
&~ \mathbf w \in \{ 0,1\}^m,
\end{array}
\tag{P0}
\label{eq: prob_sel}
\end{align}
where $\mathbf J_w \in \mathbb R^n$ is given by \eqref{eq: Jw_hat_simple}, {and $s \leq m$ is a prescribed energy budget given by the maximum number of sensors to be activated.  We recall that $n$ is the dimension of the parameter to be estimated, and $m$ is the number of sensors.}

We note that \eqref{eq: prob_sel} is a nonconvex optimization
problem due to the presence of Boolean selection variables.
Moreover, if we drop the source statistics $\boldsymbol \Sigma$ from the MSE matrix \eqref{eq: Pw_MSE} and  impose the assumption     $s \geq n$, the proposed formulation    (P0)  is then applicable for sensor selection in a non-Bayesian   framework, where the unknown parameter is estimated through the best linear unbiased estimator  \cite{karbook}.

In what follows,
we discuss  two special cases for the formulations of the sensor selection problem under two different structures of the noise covariance matrix $\mathbf R$: a) $\mathbf R$ is diagonal, and b) $\mathbf R$ has small off-diagonal entries.

\subsection{Formulation for two special cases}

When measurement noises are uncorrelated, the noise covariance matrix $\mathbf R$ becomes  diagonal. 
From \eqref{eq: Rw_sel} and \eqref{eq: Jw},   the Fisher information matrix  in the objective function of 
\eqref{eq: prob_sel}
simplifies to
\begin{align}
\mathbf J_w = &
\boldsymbol \Sigma^{-1} + \mathbf H^T \boldsymbol \Phi_w^T \boldsymbol \Phi_{w}    \mathbf R^{-1} \boldsymbol \Phi_w^T \boldsymbol \Phi_w \mathbf H \nonumber \\
= & \boldsymbol \Sigma^{-1} + \mathbf H^T \diag(\mathbf w) \mathbf R^{-1} \diag(\mathbf w)  \mathbf H   \nonumber \\
= &
\boldsymbol \Sigma^{-1} +
\sum_{i=1}^m {w_i}{R_{ii}^{-1}}  {\mathbf h_i \mathbf h_i^T },
\label{eq: JW_linear}
\end{align}
where $\mathbf h_i^T$ denotes the $i$th row of $\mathbf H$, {$R_{ii}$ denotes the $i$th diagonal entry of ${\mathbf R}$}, and
the last equality holds due to the fact that
\begin{align}
w_i^2 = w_i, ~ i =1,2,\ldots,m. 
\label{eq: w_bi_eq}
\end{align}
It is clear from  \eqref{eq: JW_linear} that each sensor contributes to Fisher information in an additive manner. As demonstrated in  \cite{SS2009} and \cite{cheleu15}, the linearity of the inverse mean squared error (Fisher information) with respect to $\mathbf w$ enables the use of convex optimization  to solve the  problem of sensor selection.

When measurement noises are weakly correlated (namely, $\mathbf R$ has small off-diagonal entries), it will be  shown in  Sec.\,\ref{sec: spe_weak} that the Fisher information matrix can be approximately expressed as
\begin{align}
\hat{\mathbf J}_w \Def \boldsymbol \Sigma^{-1} + \mathbf H^T (\mathbf w \mathbf w^T \circ \mathbf R^{-1}) \mathbf H,
\label{eq: J_weak}
\end{align}
where $\circ$ stands for the Hadamard (elementwise) product.
The problem of sensor selection  with weakly correlated noise becomes
\begin{align}
\begin{array}{ll}
\displaystyle \minimize_{\mathbf w} & ~\tr  \left ( \boldsymbol \Sigma^{-1} + \mathbf H^T (\mathbf w \mathbf w^T \circ \mathbf R^{-1}) \mathbf H \right )^{-1} \\
\st &~ \mathbf 1^T \mathbf w \leq s, \\
&~ \mathbf w \in \{ 0,1\}^m.
\end{array}
\tag{P1}
\label{eq: prob_weak}
\end{align}

Compared to the generalized  formulation \eqref{eq: prob_sel}, 
the objective function of \eqref{eq: prob_weak}  is convex with respect to the rank-one matrix $\mathbf w \mathbf w^T$. Such structure introduces computational benefits while solving   \eqref{eq: prob_weak}. {We emphasize that
 \eqref{eq: prob_weak}  has been     formulated 
in \cite{rig10_thesis,jamsimmaleu14_cf_v2,shevar14} for sensor selection with correlated noise, however,  
using this formulation, without acknowledging that it is only
valid when the correlation is weak, can lead to incorrect
sensor selection results.} 
We elaborate on the problem of sensor selection with weakly correlated noise in Sec.\,\ref{sec: spe_weak}.

\section{General Case: Proposed Optimization Methods for Sensor Selection}
\label{sec: gen_sel}

In this section, we present  two methods to solve \eqref{eq: prob_sel}: the first  is based on convex relaxation techniques, and the second is based on a greedy algorithm. First, we show that after relaxing the Boolean constraints the selection problem can be  cast as a standard semidefinite program (SDP). 
Given the solution of the relaxed \eqref{eq: prob_sel} we then
use the randomization method to generate a near-optimal  selection scheme. 
Next, we show that
given a subset of sensors, activating a new sensor  always   improves the  estimation performance. Motivated by this, 
we  present a greedy algorithm that scales gracefully with the problem size
to obtain locally optimal solutions of \eqref{eq: prob_sel}.

\subsection{Convex relaxation}
Substituting the expression of Fisher information \eqref{eq: Jw_hat_simple} into
problem \eqref{eq: prob_sel}, we obtain
\begin{align}
\hspace*{-0.6in}\begin{array}{ll}
\displaystyle \minimize_{\mathbf w} 
& ~\tr 
\left (
\mathbf C - \mathbf B^T \left 
(\mathbf S^{-1} + a^{-1} \diag(\mathbf w)
\right
) ^{-1}  \mathbf B
\right)^{-1}
\\
\st &~ \mathbf 1^T \mathbf w \leq s, \\
&~ \mathbf w \in \{ 0,1\}^m,
\end{array}
\hspace*{-0.4in}\label{eq: prob_sel_eq_simple}
\end{align}
where for notational simplicity we have defined
$\mathbf C \Def \boldsymbol \Sigma^{-1} + \mathbf H^T \mathbf S^{-1} \mathbf H$ and $\mathbf B \Def \mathbf S^{-1} \mathbf H$.

Problem \eqref{eq: prob_sel_eq_simple} can be equivalently transformed to \cite{Boyd2004_bk}
\begin{align}
\hspace*{-0.6in}\begin{array}{ll}
\displaystyle \minimize_{\mathbf w, \mathbf Z} 
& \tr \left(
\mathbf Z
\right )\\
\st 
& \mathbf C - \mathbf B^T \left 
(\mathbf S^{-1} + a^{-1} \diag(\mathbf w)
\right
) ^{-1}  \mathbf B \succeq \mathbf Z^{-1},\\
& \mathbf 1^T \mathbf w \leq s, \\
& \mathbf w \in \{ 0,1\}^m,
\end{array}
\hspace*{-0.4in}\label{eq: prob_sel_eq_epi}
\end{align}
where  $\mathbf Z \in \mathbb S^n$ is an auxiliary variable,  {$\mathbb S^n$ represents the set of $n \times n$ symmetric matrices},   and the notation $ \mathbf X \succeq  \mathbf Y$ (or $ \mathbf X \preceq  \mathbf Y$) indicates that the matrix $\mathbf X - \mathbf Y$ (or $\mathbf Y - \mathbf X$) is positive semidefinite.
The first inequality constraint in \eqref{eq: prob_sel_eq_epi} is obtained from
\[
\left ( \mathbf C - \mathbf B^T \left 
(\mathbf S^{-1} + a^{-1} \diag(\mathbf w)
\right
) ^{-1}  \mathbf B \right )^{-1} \preceq \mathbf Z,
\]
which implicitly adds the additional constraint $\mathbf Z \succeq 0$, since the left hand side of the above inequality is the inverse of the Fisher information matrix.

We further introduce another auxiliary variable $\mathbf V \in \mathbb S^n$ such that
the first matrix inequality of \eqref{eq: prob_sel_eq_epi} is  expressed as
\begin{align}
&\mathbf C - \mathbf V \succeq \mathbf Z^{-1}, 
\label{eq: CVZ} 
\end{align}
and
\begin{align}
\mathbf V \succeq \mathbf B^T \left (\mathbf S^{-1} + a^{-1} \diag(\mathbf w)  \right) ^{-1} \mathbf B. \label{eq: V_newvar_ineq}
\end{align}
Note that the minimization of  $\tr(\mathbf Z)$ with inequalities \eqref{eq: CVZ} and \eqref{eq: V_newvar_ineq} would force 
the variable $\mathbf V$ to achieve its lower bound.
In other words, problem \eqref{eq: prob_sel_eq_epi} is equivalent to the problem in which the inequality constraint in \eqref{eq: prob_sel_eq_epi} is replaced by the two inequalities \eqref{eq: CVZ} and \eqref{eq: V_newvar_ineq}.
Finally, employing the Schur complement,  the inequalities \eqref{eq: CVZ} and \eqref{eq: V_newvar_ineq} can be rewritten as the following linear matrix inequalities (LMIs)
\begin{align}
\begin{bmatrix}
 \mathbf C-\mathbf V & \mathbf I \\
\mathbf I & \mathbf Z
\end{bmatrix} \succeq 0,~
\begin{bmatrix}
\mathbf V    & \mathbf B^T\\
\mathbf B &\mathbf S^{-1} + a^{-1} \diag(\mathbf w)
\end{bmatrix} \succeq 0.
\label{eq: Schur2}
\end{align}

Substituting \eqref{eq: Schur2} into \eqref{eq: prob_sel_eq_epi},
the sensor selection problem becomes
\begin{align}
\hspace*{-0.6in}\begin{array}{ll}
\displaystyle \minimize_{\mathbf w, \mathbf Z, \mathbf V} 
& ~\tr \left(
\mathbf Z
\right )\\
\st 
& ~\text{LMIs in \eqref{eq: Schur2}},\\
&~ \mathbf 1^T \mathbf w \leq s, \\
&~ \mathbf w \in \{ 0,1\}^m.
\end{array}
\hspace*{-0.4in}\label{eq: prob_sel_eq_LMI}
\end{align}
Problem \eqref{eq: prob_sel_eq_LMI} has the form of an SDP except  for the last Boolean constraints.
As shown in \cite{SS2009}, one possibility is to relax each Boolean variable to its convex hull to obtain $\mathbf w \in [0,1]^m$.
In this case, we can choose $s$ active sensors given by
the first $s$ largest entries of the solution   of  the relaxed problem, or employ a randomized rounding algorithm  \cite[Algorithm\,3]{cheleu15}  to generate a Boolean selection vector.

Rather than directly relaxing   Boolean selection variables to continuous variables, 
we can use semidefinite relaxation (SDR) \cite{luomasoyezha10} --- referred to problems in which the relaxation of a rank constraint leads to an SDP --- to better overcome the difficulties posed by the nonconvex constraints of \eqref{eq: prob_sel_eq_LMI}. 
The Boolean constraint \eqref{eq: w_bi_eq} on the entries of $\mathbf w$ can be enforced by
\begin{align}
\diag(\mathbf w \mathbf w^T) = \mathbf w, \label{eq: boo_w}
\end{align}
where, with an abuse of notation,  $\diag( \cdot )$ 
returns in vector form the diagonal entries of its matrix argument.
By introducing an auxiliary variable $\mathbf W$ together with the rank-one constraint 
\begin{align}
\mathbf W = \mathbf w \mathbf w^T, \label{eq: rank1_cons}
\end{align}
the energy and Boolean constraints in \eqref{eq: prob_sel_eq_LMI} can be expressed as
\begin{align}
\tr(\mathbf W) \leq s,~  ~\diag(\mathbf W) = \mathbf w. \label{eq: Xw_ene_boo}
\end{align}
After relaxing  the  (nonconvex) rank-one constraint \eqref{eq: rank1_cons} to $\mathbf W \succeq \mathbf w \mathbf w^T$, we reach the  SDP
\begin{align}
\hspace*{-0.6in}\begin{array}{ll}
\displaystyle \minimize_{\mathbf w, \mathbf W, \mathbf Z, \mathbf V} 
& ~\tr \left(
\mathbf Z
\right )\\
\st 
& ~\text{LMIs in \eqref{eq: Schur2}},\\
&~ \tr(\mathbf W) \leq s, \\
&~ \diag(\mathbf W) = \mathbf w, \\
&~ \begin{bmatrix}
\mathbf W & \mathbf w \\
\mathbf w^T & 1
\end{bmatrix} \succeq 0,
\end{array}
\hspace*{-0.4in}\label{eq: prob_sel_eq_LMI_SDR}
\end{align}
where the last inequality is derived through the application of a Schur complement to $\mathbf W \succeq \mathbf w \mathbf w^T$.

We can use an interior-point algorithm to solve the SDP \eqref{eq: prob_sel_eq_LMI_SDR}. 
In practice, if  the dimension of the unknown parameter  vector is much less than the number of sensors, the computational complexity of SDP 
is roughly given by $O(m^{4.5})$ \cite{Interior_bk}.  
Once the SDP \eqref{eq: prob_sel_eq_LMI_SDR} is solved, we  employ a randomization method to generate a near-optimal sensor selection scheme, where the effectiveness of the randomization method has been shown in our extensive numerical experiments. We refer the readers to \cite{luomasoyezha10} for more details on the motivation and benefits of  randomization used in SDR.
The aforementioned procedure is summarized in Algorithm\,1,
 which   includes the randomization procedure described in Algorithm\,2.

\begin{algorithm}
\caption{SDR with randomization for sensor selection}
\begin{algorithmic}[1]
\Require prior information $\boldsymbol \Sigma$, $\mathbf R = a \mathbf I  + \mathbf S$ as in \eqref{eq: R_new}, observation matrix $\mathbf H$ and energy budget $s$
\State  solve the   SDP \eqref{eq: prob_sel_eq_LMI_SDR} and    obtain solution $(\mathbf w, \mathbf W)$ 
\State call Algorithm\,2 for Boolean solution.
\end{algorithmic}
\end{algorithm}

\begin{algorithm}
\caption{Randomization method  \cite{luomasoyezha10}}
\begin{algorithmic}[1]
\Require solution pair $(\mathbf w, \mathbf W)$ from the SDP \eqref{eq: prob_sel_eq_LMI_SDR}
\For {$l = 1, 2, \ldots, N$}
\State  pick a random number $\boldsymbol \xi^{(l)} \sim \mathcal N(\mathbf w, \mathbf W - \mathbf w \mathbf w^T)$
\State map $\boldsymbol \xi^{(l)}$ to a sub-optimal sensor selection scheme $\mathbf w^{(l)}$
\begin{align*}
  w_j^{(l)} = \left \{ \begin{array}{ll}
 1 &    \xi_j^{(l)} \geq [\boldsymbol \xi^{(l)}]_s \\
0 & \text{otherwise} ,
\end{array} ~ j =1,2,\ldots,m,
\right.
\end{align*}
\hspace*{0.21in}where $  w_j^{(l)}$ is the $j$th element of $\mathbf w^{(l)}$, and
$ [\boldsymbol \xi^{(l)}]_s $ 
\hspace*{0.21in}denotes the $s$th largest entry of 
$\boldsymbol \xi^{(l)}
$
\EndFor
\State choose a vector in $\{ \mathbf w^{(l)}\}_{l=1}^N$ 
which yields the smallest  objective value of \eqref{eq: prob_sel_eq_simple}. 
\end{algorithmic}
\end{algorithm}

\subsection{Greedy algorithm}

We begin by showing in Proposition\,\ref{prop: FIM}  that 
even in the presence of correlated measurement noise, the  Fisher information increases if an inactive sensor is made active.

\begin{myprop}
\label{prop: FIM}
If $\mathbf w$ and {$\tilde{\mathbf w}$} represent two sensor selection vectors, where $w_i = \tilde w_i$ for $i \in \{ 1,2,\ldots,m \} \setminus \{ j\}$,  $w_j = 0$ and $\tilde w_j = 1$, then the resulting Fisher information matrix satisfies
$\mathbf J_{\tilde w} \succeq \mathbf J_w$.
More precisely, {
\begin{align}
\mathbf J_{\tilde w} - \mathbf J_{w} = c_j \boldsymbol \alpha_j \boldsymbol \alpha_j^T,
\label{eq: rank_one_FIM}
\end{align}
and
\begin{align}
\tr(\mathbf J_w^{-1}) - \tr(\mathbf J_{\tilde w}^{-1})
= \frac{c_j \boldsymbol \alpha_j^T \mathbf J_w^{-2} \boldsymbol \alpha_j}{1 +c_j \boldsymbol \alpha_j \mathbf J_w^{-1} \boldsymbol \alpha_j } \geq 0,
\label{eq: trP_vw}
\end{align}
where
$c_j$ is a positive scalar given by
\begin{align}
c_j = \left \{ 
\begin{array}{ll}
R_{jj}^{-1} & \mathbf w = \mathbf 0 \\
(R_{jj} - \mathbf r_j^T \mathbf R_w^{-1} \mathbf r_j)^{-1} & \text{otherwise},
\end{array}
\right .
\label{eq: cj_prop2}
\end{align}
and 
\begin{align}
\boldsymbol \alpha_j = \left \{ 
\begin{array}{ll}
\mathbf h_j & \mathbf w = \mathbf 0 \\
\mathbf H^T \boldsymbol \Phi_w^T \mathbf R_w^{-1} \mathbf r_j - \mathbf h_j & \text{otherwise}.
\end{array}
\right .
\label{eq: alphaj_prop2}
\end{align}
In \eqref{eq: cj_prop2}-\eqref{eq: alphaj_prop2},
$R_{jj}$ is the $j$th diagonal entry of $\mathbf R$, {$\mathbf r_j$ represents the covariance vector between the measurement noise of the $j$th sensor and that of the active sensors in $\mathbf w$,}
 $\mathbf h_j^T$ is the $j$th row of $\mathbf H$,  $\boldsymbol \Phi_w$ and $\mathbf R_w$  are given by   \eqref{eq: meas_Phiw} and \eqref{eq: Rw_sel}, respectively. }
\end{myprop}
\textbf{Proof}: See Appendix\,\ref{app:prop1}. \hfill $\blacksquare$

It is clear from  \eqref{eq: rank_one_FIM} that when an inactive sensor is made active, {the increase in Fisher information leads to an information gain in terms of the rank-one matrix given by \eqref{eq: rank_one_FIM}}. Such a  phenomenon was also discovered in  the calculation of sensor utility for adaptive signal estimation \cite{berszuruc12} and leader selection in stochastically forced consensus networks \cite{linfarjov14}.
Since
activating a new sensor does not degrade the estimation performance,  the inequality (energy) constraint in \eqref{eq: prob_sel} can be reformulated as an equality constraint.

In a greedy algorithm, we iteratively select a new sensor which gives the largest  {performance improvement} until the energy constraint is satisfied with equality. {The greedy algorithm is attractive due to its simplicity}, and has been employed in a variety of applications \cite{berszuruc12,linfarjov14,shabanvik10}. In particular, 
a greedy algorithm was  proposed in \cite{shabanvik10} for sensor selection   under the assumption of   uncorrelated measurement noise. 
We generalize the framework of \cite{shabanvik10} by taking into account  noise correlation.
 {Clearly, in each iteration of the greedy algorithm,
 the newly activated sensor is the one that maximizes the performance improvement characterized by $\tr(\mathbf J_w^{-1}) -  \tr(\mathbf J_{\tilde w}^{-1})$  in \eqref{eq: trP_vw}.
We summarize the greedy algorithm in Algorithm\,3.
}

\begin{algorithm}
\caption{Greedy algorithm for  sensor selection}
\begin{algorithmic}[1]
\Require $\mathbf w = \mathbf 0$, $ {\mathcal I}  = \{ 1,2, \ldots,m\}$ and $\mathbf J_w = \boldsymbol \Sigma^{-1}$
\For {$l = 1, 2, \ldots, s$}
\State given $\mathbf w$, {enumerate all the inactive sensors in  $ {\mathcal I}$ to }
\hspace*{0.21in}determine  $j \in   {\mathcal I}$ such that  
$\tr(\mathbf J_w^{-1}) -  \tr(\mathbf J_{\tilde w}^{-1})$ in \eqref{eq: trP_vw} 
\hspace*{0.21in}is maximized
\State update $\mathbf w$ by setting $w_j = 1$, and update  $\mathbf J_w$ by adding 
\hspace*{0.21in}$c_j \boldsymbol \alpha_j \boldsymbol \alpha_j^T$ in \eqref{eq: rank_one_FIM}
\State remove $j$ from ${\mathcal I} $.
\EndFor
\end{algorithmic}
\end{algorithm}

In Step\,2 of Algorithm\,3, we search $O(m)$ sensors to achieve the largest performance improvement. 
In \eqref{eq: trP_vw}, the computation of $\mathbf J_w^{-1}$ incurs a 
  complexity of $O(n^{2.373})$ \cite{Williams12}.
Since Algorithm\,3 terminates after $s$ iterations,  its overall complexity  is  given by $O(sm + sn^{2.373})$, where at each iteration, 
the calculation of  $\mathbf J_w^{-1}$ is independent of the search for the new active sensor. If  the dimension of $\mathbf x$    is much less than the number of sensors, the complexity of Algorithm\,3 reduces to $O(sm)$. {Our extensive numerical  experiments  show that the greedy algorithm is able to yield good locally optimal sensor selection schemes.}

\section{Special Case: Sensor Selection with Weak Noise Correlation}
\label{sec: spe_weak}
{In this section, we   show that   
the existing sensor selection model in \cite{rig10_thesis,jamsimmaleu14_cf_v2,shevar14} is invalid   for an arbitrary noise covariance matrix.
We 
establish that    in contrast to the approach proposed in this paper, the existing model in \cite{rig10_thesis,jamsimmaleu14_cf_v2,shevar14} is  only valid   when measurement noises are weakly correlated.
In this scenario, the proposed   sensor selection problem given by \eqref{eq: prob_sel} would simplify to \eqref{eq: prob_weak}.}
Moreover, if the trace of  the Fisher information   matrix (also known as information gain defined in \cite{shevar14})  is adopted as the performance measure for sensor selection,  
we show that the resulting optimization problem can be  cast as a special problem of maximizing a convex quadratic
function over a bounded polyhedron. 

\subsection{Drawbacks  of existing  formulation}
In \cite{rig10_thesis,jamsimmaleu14_cf_v2,shevar14}, 
several variations of sensor selection problems with correlated noise have been
studied, based on whether the quantity to be estimated is a random parameter or a random process, and whether the cost function is energy or estimation error.
The common feature in \cite{rig10_thesis,jamsimmaleu14_cf_v2,shevar14} is that  the  information matrix was  approximated by \eqref{eq: J_weak};  we repeat equation \eqref{eq: J_weak} here for 
convenience
\begin{align}
\hat{\mathbf J}_w = \boldsymbol \Sigma^{-1} + \mathbf H^T (\mathbf w \mathbf w^T \circ \mathbf R^{-1}) \mathbf H.
\label{eq: J_weak_repeat}
\end{align}
Compared to our   formulation \eqref{eq: Jw},
 the noise covariance matrix appearing in \eqref{eq: J_weak_repeat} is independent of the sensor selection variables. In fact,
$\hat{\mathbf J}_w$ 
can be thought of as Fisher information under the measurement model
\begin{align}
\mathbf y = \boldsymbol \Phi_w \mathbf H \mathbf x + \mathbf v,
\label{eq: meas_exist}
\end{align}
where $\boldsymbol \Phi_w$ was defined in \eqref{eq: meas_Phiw}. 
Different from \eqref{eq: meas_Phiw}, 
{the noise from the unselected sensors is spread across the selected sensors. As a result, the measurement model \eqref{eq: meas_exist} yields
$y_i = v_i$ if  the $i$th sensor is inactive. This contradicts  the fact that an inactive sensor should keep silent and thus have no effect on the estimation task.
}

The Fisher information in \eqref{eq: J_weak_repeat}
can also be interpreted as  \cite[Sec.\,3]{rig10_thesis}
\begin{align}
\hat {\mathbf J}_w &= \boldsymbol \Sigma^{-1} + \sum_{i,j \in \mathcal S} \bar{R}_{ij}\mathbf h_i \mathbf h_j^T, \nonumber \\
& = \boldsymbol \Sigma^{-1} +  \mathbf H^T \boldsymbol \Phi_w^T  (\boldsymbol \Phi_w  \mathbf R^{-1} \boldsymbol \Phi_w^T) \boldsymbol \Phi_w \mathbf H,
\label{eq: J_weak_thesis}
\end{align}
where $\mathcal S$ is the set of selected sensors, and {$\bar R_{ij}$ denotes the $(i,j)$th entry of $\mathbf R^{-1}$}. 
In \eqref{eq: J_weak_thesis}, $\mathbf R^{-1}$ is computed first and then truncated according to the sensor selection scheme. This is an incorrect way of modeling the noise covariance matrix for active sensors, since the matrix $\mathbf R$ should be truncated first and then inverted  
 as demonstrated in \eqref{eq: Jw}.

Both of the interpretations \eqref{eq: meas_exist} and \eqref{eq: J_weak_thesis} indicate that 
the existing formulation in \cite{rig10_thesis,jamsimmaleu14_cf_v2,shevar14} is inaccurate for modeling the problem of sensor selection with correlated noise.
{A natural question that arises from the preceding discussion is whether there exist  a condition  that ensures  the validity of the Fisher information matrix \eqref{eq: J_weak_repeat} as presented in \cite{rig10_thesis,jamsimmaleu14_cf_v2,shevar14}? We will show in the next section that the formulation reported in  \cite{rig10_thesis,jamsimmaleu14_cf_v2,shevar14}   becomes valid   only when sensor selection is restricted to the weak noise correlation regime.  
}

\subsection{Validity of existing  formulation: weak correlation}

{We consider the scenario of weakly correlated noise, in which the noise covariance matrix $\mathbf R$ has small off-diagonal entries, namely, noises are weakly correlated across the sensors. For ease of representation, we express the   noise covariance matrix    as
\begin{align}
\mathbf R = \boldsymbol \Lambda + \epsilon \boldsymbol \Upsilon,
\label{eq: R_weak}
\end{align}
where $\boldsymbol \Lambda$ is a diagonal matrix which consists of the diagonal entries of $\mathbf R$,   $\epsilon \boldsymbol \Upsilon$ is a symmetric matrix whose diagonal entries are zero and off-diagonal entries correspond to those  of $\mathbf R$,   the parameter $\epsilon$ is introduced to govern  the strength of noise correlation across the sensors, and $\boldsymbol \Lambda$ and $\boldsymbol \Upsilon$ are independent of  $\epsilon$.  
Clearly,  the covariance   of weakly correlated noises can be described by \eqref{eq: R_weak} for some small value of $\epsilon$ since $\boldsymbol \Upsilon$ is $\epsilon$-independent. As  $\epsilon \to 0$, the off-diagonal 
 entries of $\mathbf R$ are   forced to go to zero. 

Proposition\,\ref{prop: weak}  below shows that  the  correct expression \eqref{eq: Jw} of  Fisher information  is equal  to  the  expression  \eqref{eq: J_weak_repeat}, as presented in \cite{rig10_thesis,jamsimmaleu14_cf_v2,shevar14},  
 up to first order in $\epsilon$ as $\epsilon \to 0$.

\begin{myprop}
\label{prop: weak}
If measurement noises are weakly correlated and $\mathbf R = \boldsymbol \Lambda + \epsilon \boldsymbol \Upsilon$, 
then the Fisher information matrix \eqref{eq: Jw} can be expressed as
\[
\mathbf J_w = \hat{\mathbf J}_w + O(\epsilon^2)  \quad \text{as $\epsilon \to 0$},
\]
where $ \hat{\mathbf J}_w$ is given by \eqref{eq: J_weak_repeat}.
\end{myprop}
\textbf{Proof:} See Appendix\,\ref{prob: Jweak}. \hfill $\blacksquare$
}

It is clear from Proposition\,\ref{prop: weak} that \eqref{eq: prob_weak} is  valid only when the noise correlation is weak.
Proceeding with the same logic as in the previous section for the introduction of constraints \eqref{eq: rank1_cons}-\eqref{eq: Xw_ene_boo},
{we relax \eqref{eq: prob_weak}  to the SDP 
\begin{align}
\begin{array}{ll}
\displaystyle \minimize_{\mathbf w, \mathbf W, \mathbf Z} &~ \tr (\mathbf Z) \\
\st &~ \begin{bmatrix}
 \boldsymbol \Sigma^{-1} + \mathbf H^T (\mathbf W \circ \mathbf R^{-1}) \mathbf H & \mathbf I \\
\mathbf I & \mathbf Z
\end{bmatrix} \succeq 0 , \\
&~ \tr(\mathbf W) \leq s,  ~ \diag(\mathbf W) = \mathbf w, \\
& ~ \begin{bmatrix}
\mathbf W & \mathbf w \\
\mathbf w^T & 1
\end{bmatrix} \succeq 0,
\end{array}
\label{eq: prob_weak_X_relax}
\end{align}
where $\mathbf Z \in \mathbb S^n$ is an auxiliary optimization variable.}
Given the solution pair $(\mathbf w, \mathbf W)$ of problem \eqref{eq: prob_weak_X_relax}, we can use the randomization method  in Algorithm\,2 to construct a near-optimal sensor selection scheme.
{The computational complexity of solving problem \eqref{eq: prob_weak_X_relax} is close to that of solving the SDP \eqref{eq: prob_sel_eq_LMI_SDR}. However, as will be evident later, the   sensor selection problem with  weakly correlated noise can be further simplified if
 the trace of  the Fisher information   matrix   is used as the performance measure.
In this scenario, 
the obtained problem structure   enables the use of more computationally inexpensive algorithms, e.g., bilinear programing, to solve the sensor selection problem. 
}

\subsection{Sensor selection by maximizing   trace of Fisher information}

Instead of minimizing the estimation error,  the trace of  Fisher information {(so-called T-optimality \cite{F1994})}   also has been used     as a  performance metric in problems of sensor selection \cite{shevar14, sheliuvar14,TM2011}. 
{According to  \cite[Lemma\,1]{fanli09_tsp}, the trace of Fisher information constitutes a lower bound to the trace of {error covariance matrix given by $\mathbf J_w^{-1}$ in \eqref{eq: Jw}}. That is,
\begin{align}
\tr(\mathbf J_w^{-1}) \geq \frac{n^2}{\tr(\mathbf J_w)}. \label{eq: trJ_trP}
\end{align}
Motivated by \eqref{eq: trJ_trP} and the generalized information gain used in  \cite{shevar14}, we propose  to minimize the lower bound of the objective function in 
{\eqref{eq: prob_weak}}, which leads to the     problem
\begin{align}
\begin{array}{ll}
\displaystyle \maximize_{\mathbf w} & ~ \tr  \left ( \boldsymbol \Sigma^{-1} + \mathbf H^T (\mathbf w \mathbf w^T \circ \mathbf R^{-1}) \mathbf H \right )\\
\st &~ \mathbf 1^T \mathbf w \leq s, \\
&~ \mathbf w \in \{ 0,1\}^m.
\end{array}
\hspace*{-0.2in}\label{eq: prob_weak_trJ}
\tag{P2}
\end{align}

It is worth mentioning that the  sensor  selection scheme obtained from \eqref{eq: prob_weak_trJ}   may not be optimal in the  MMSE sense. However, the trace operator is linear and   introduces computational benefits in optimization. 
Reference \cite{shevar14}  has shown that  \eqref{eq: prob_weak_trJ} is  not convex   even if 
Boolean selection variables are relaxed.
However, there is no theoretical
justification and analysis provided in \cite{shevar14}  on the
problem structure.  
In what follows, we demonstrate that the Boolean constraint in \eqref{eq: prob_weak_trJ} can be replaced by its convex hull $\mathbf w \in [0,1]^m$ without loss of performance, to obtain an equivalent optimization problem.

\begin{myprop}
\label{prop: trJ}
 \eqref{eq: prob_weak_trJ} is equivalent to 
 \begin{align}
\begin{array}{ll}
\displaystyle \maximize_{\mathbf w} & ~  \mathbf w^T \boldsymbol \Omega \mathbf w\\
\st &~ \mathbf 1^T \mathbf w \leq s, \\
&~   \mathbf w \in [  0,   1]^m,
\end{array}
\label{eq: prob_weak_w_relax}
\end{align}
where $\boldsymbol \Omega$ is a positive semidefinite matrix given by
$
 \mathbf A (\mathbf R^{-1} \otimes \mathbf I_{n }) \mathbf A^T
$, $\otimes$
denotes the Kronecker product,
$
\mathbf A \in \mathbb R^{m \times mn}
$ is a block-diagonal matrix whose diagonal blocks are given by $\{ \mathbf h_i^T \}_{i=1}^m$,
and
$
\mathbf h_i^T
$ denotes the $i$th  row of  the measurement matrix $\mathbf H$.
\end{myprop}
\textbf{Proof:}
{See Appendix\,\ref{app: prop_trJ}.}\hfill $\blacksquare$

It is clear from Proposition\,\ref{prop: trJ} that  \eqref{eq: prob_weak_trJ} eventually approaches the problem of maximizing a convex quadratic function over a bounded polyhedron.
It is known \cite{par91} that finding a globally optimal solution of   \eqref{eq: prob_weak_w_relax} is NP-hard. 
Therefore, we resort to local optimization methods, such as bilinear programming and SDR, to solve   problem \eqref{eq: prob_weak_w_relax}. 
To be specific, bilinear programming is a special case of alternating convex optimization, where at each    iteration we solve two linear programs. 
Since bilinear programming is based on linear programming, it scales gracefully with problem size but with a possibility of only finding local optima.
If we rewrite  the constraints of problem \eqref{eq: prob_weak_w_relax} as quadratic forms in $\mathbf w$,  \eqref{eq: prob_weak_trJ} can be further transformed into a nonconvex \textit{homogeneous} quadratically constrained
quadratic program (QCQP), which refers to a QCQP without involving linear terms of optimization variables.
In this scenario, SDR can be applied to solve the problem. 
Compared to the application of SDR in \eqref{eq: prob_weak_X_relax}, 
the  homogeneous QCQP leads to an SDP  with a smaller problem size. 
We refer the readers to \cite[Sec.\,V]{liumasfarvar15_cf} and 
\cite[Sec.\,V]{shevar14} for more details on the application of bilinear programming and SDR.
}

\section{Non-myopic Sensor Scheduling}
\label{sec: non_mop}
In this section, we extend the sensor selection framework  
with correlated noise to the problem of non-myopic sensor
scheduling, which determines sensor activations for multiple
future time steps. Since the Fisher information matrices at
consecutive time steps are coupled with each other, expressing
them in a closed form with respect to the sensor selection
variables becomes intractable. Therefore, we employ a greedy algorithm to seek  locally optimal solutions of the non-myopic sensor scheduling problem.

Consider a discrete-time dynamical system
\begin{align}
\mathbf x_{t+1} & = \mathbf F_t \mathbf x_{t}+ \mathbf u_{t} \label{eq: state_gen}\\
\mathbf y_t & = \mathbf H_t \mathbf x_t + \mathbf v_t,
\label{eq: meas_gen}
\end{align}
where $\mathbf x_t \in \mathbb R^n$ is the target state at time $t$, $\mathbf y_t \in \mathbb R^m$ is  the measurement vector whose $i$th entry corresponds to a scalar observation from the $i$th sensor at time $t$, {$\mathbf F_t$ is the state transition matrix from time $t$ to time $t+1$,}
 and $\mathbf H_t$  denotes the  observation matrix at time  $t$. The inputs $\mathbf u_t$ and $\mathbf v_t$
 are white, Gaussian, zero-mean random vectors with covariance matrices $\mathbf Q$ and $\mathbf R$, respectively. 
We note that the covariance matrix $\mathbf R$ may not be diagonal, since the 
noises experienced by different sensors could be spatially {correlated}. 
We also remark that although the dynamical system \eqref{eq: state_gen}-\eqref{eq: meas_gen} is assumed to be linear, it will be evident later that the proposed sensor scheduling framework is also applicable to   non-linear dynamical systems.

{The PDF of the initial state $\mathbf x_0$ at time step $t_0$ is assumed to be Gaussian  with mean $\hat{\mathbf x}_0$ and covariance matrix $\hat{\mathbf P}_0$,}
where   $\hat{\mathbf x}_{0}$ and $\hat{\mathbf P}_{0}$  are estimates of the initial state and error covariance from the previous measurements obtained using filtering algorithms, such as  a particle filter or  a Kalman filter \cite{MRV2012, masfarvar12}.
At  time step $t_0$, we aim to 
find the optimal sensor schedule over the next $\tau$ time steps $t_0+1, t_0 +2, \ldots, t_0 + \tau$.
Hereafter, for notational simplicity, we assume $t_0 = 0$.
The sensor schedule can be represented by a vector of binary variables
\begin{align}
\mathbf w = [\mathbf w_{1}^T, \mathbf w_{2}^T, \ldots, \mathbf w_{\tau}^T]^T \in \{ 0,1 \}^{\tau m}, \label{eq: w_schedule}
\end{align}
where
$
\mathbf w_{t} = [ w_{t,1}, w_{t,2}, \ldots, w_{t,m}]^T
$ characterizes the sensor schedule at time  $ 1 \leq t \leq \tau$.
In what follows, we assume that $\tau > 1$. If  $\tau = 1$,  the non-myopic sensor scheduling problem reduces to   the sensor selection problem for one snapshot or the so-called myopic scheduling problem. {This case has been  studied in the previous sections.}

In the context of state tracking \cite{TMN1998,HK_bk},  {the} Fisher information matrix   has the following recursive form
\begin{align}
\mathbf J_t & = (\mathbf Q + \mathbf F_{t-1} \mathbf J_{t-1}^{-1} \mathbf F_{t-1}^T)^{-1} + \mathbf G_t \label{eq: FIM_recursive}\\
\mathbf G_t &= \mathbf H_t^T \boldsymbol \Phi_{w_t}^T ( \boldsymbol \Phi_{w_t}  \mathbf R \boldsymbol \Phi_{w_t}^T)^{-1} \boldsymbol \Phi_{w_t} \mathbf H_t,
\label{eq: Gt}
\end{align}
for {$t = 1, 2, \ldots,  \tau$}, where $\mathbf J_t$ denotes the Fisher information at time {$t$}, $\mathbf G_t $ denotes the part of Fisher information matrix which incorporates the updated measurement,
and 
$
\boldsymbol \Phi_{w_t}
$ 
is a submatrix of $\diag(\mathbf w_t)$ where all {the} rows
corresponding to the unselected sensors are removed.
 It is clear from \eqref{eq: PhiSPhi} that  the term involving $ \boldsymbol \Phi_{w_t}$ in \eqref{eq: Gt}  can be further expressed as an explicit form with respect to $\mathbf w_t$.

\begin{remark}
\textit{
In case of non-linear measurement models, the term $\mathbf G_t$ in the Fisher information matrix becomes
\begin{align*}
\mathbf G_t &= \mathbf E_{\mathbf x_t} [ (\nabla_{\mathbf x_t^T} \mathbf h)^T \boldsymbol \Phi_{w_t}^T ( \boldsymbol \Phi_{w_t}  \mathbf R \boldsymbol \Phi_{w_t}^T)^{-1} \boldsymbol \Phi_{w_t} (\nabla_{\mathbf x_t^T} \mathbf h)],
\end{align*}
where $\mathbf h(\cdot)$ is a {nonlinear} measurement function, and $\nabla_{\mathbf x_t^T} \mathbf h$ is the Jacobian matrix of $\mathbf h$ with respect to {$\mathbf x_t$}. In this equation, the expectation with respect to $\mathbf x_t$ is commonly calculated with the help of the prediction state $\hat {\mathbf x}_t  \Def \mathbf F_{t-1} \mathbf F_{t-2}   \cdots   \mathbf F_0   \hat{\mathbf x}_0 $ \cite{masfarvar12,cheleu14}. 
To be concrete, we approximate the PDF of $\mathbf x_t$ with $p(\mathbf x_t) = \delta (\mathbf x_t - \hat {\mathbf x}_t)$, where $\delta(\cdot)$ is  a  $\delta$-function. The matrix $\mathbf G_t$   is then given by
\begin{align}
\mathbf G_t  =   \hat{\mathbf H}_t^T \boldsymbol \Phi_{w_t}^T ( \boldsymbol \Phi_{w_t}  \mathbf R \boldsymbol \Phi_{w_t}^T)^{-1} \boldsymbol \Phi_{w_t}\hat{\mathbf H}_t,
\label{eq: Gt_particle}
\end{align}
where  $\hat{\mathbf H}_t \Def \nabla_{\mathbf x_t^T} \mathbf h(\hat {\mathbf x}_t)$.
}
\label{Remark: nonlinear}
\end{remark}

We note that the Fisher information matrices  at consecutive time steps are coupled with each other due to the recursive structure in \eqref{eq: FIM_recursive}. Therefore, $\mathbf J_t$ is a function of {all} selection variables {$\{  \mathbf w_{ k} \}_{k = 1}^{ t}$}. The recursive structure makes the closed form of  Fisher information intractable. This is in sharp contrast with the problem of myopic sensor selection, where  {expressing} {the} Fisher information {matrix} {in a closed form}  is {possible}. 

We now pose the non-myopic sensor scheduling problem
\begin{subequations}
\begin{align}
\displaystyle \minimize_{\mathbf w}  & ~ \frac{1}{\tau}\sum_{t = 1}^{\tau} \tr(\mathbf J_t^{-1})  \nonumber \\
\st & ~ \mathbf 1^T \mathbf w \leq s,\label{eq: cum_ene} \\ 
&~ \textstyle{ \sum_{t=1}^{\tau} w_{t,i} \leq s_i},~ i =1,2,\ldots,m,\label{eq: ini_ene}\\
& ~\mathbf w \in \{ 0,1\}^{m\tau}, \nonumber
\end{align}
\label{eq: prob_sch}
\end{subequations}
\hspace*{-0.05in}where $\mathbf J_t$ is determined by \eqref{eq: FIM_recursive}-\eqref{eq: Gt}, 
the \textit{cumulative} energy constraint \eqref{eq: cum_ene} restricts the total number of   activations for all  sensors over the entire time horizon, 
and
 the  \textit{individual} energy constraint 
\eqref{eq: ini_ene}  implies that
the $i$th sensor can report at most $s_i$
measurements over $\tau$ time steps.

To solve problem \eqref{eq: prob_sch} in a numerically efficient manner, 
we employ a greedy algorithm {that iteratively activates one sensor at a time   until the energy constraints are satisfied with equality}. The proposed greedy algorithm can be viewed as   a generalization  of Algorithm\,3 by incorporating the {length of the time horizon} and   {individual energy constraints}.

We elaborate on the greedy algorithm. In the initial step,
we assume $\mathbf w = \mathbf 0$ and split the set of indices of $\mathbf w$ into $m$
subsets $\{ \mathcal I_i\}_{i=1}^m$, where
we use the entries of the set $\mathcal I_i$ to keep track of all the time instants at which the $i$th sensor is inactive. The set $\mathcal I_i$ is initially given by
$\{ i, i +m, \ldots, i +(\tau -1)m\}$ for $i=1,2,\ldots,m$.
{There exists a one-to-one correspondence between an index  $j \in \mathcal I_i$ and a time instant  $t \in \{ 1,2,\ldots,\tau\}$ at which the $i$th sensor can be scheduled, where $j = i+ (t-1)m$.
At every iteration of the  greedy optimization algorithm, 
we update  $\mathcal I_i $ for $i=1,2,\ldots,m$ such that it only contains
 indices of zero entries of $\mathbf w$. The quantity $\tau - | \mathcal I_i |$ gives the number of times that the $i$th sensor has been used, where $|\cdot|$ denotes the cardinality of a set. {The condition $\tau - | \mathcal I_i | \geq s_i$ indicates a violation of the individual energy constraint.}
Note that the union $\{  {\mathcal I}_1 \cup {\mathcal I}_2 \cup \ldots \cup {\mathcal I}_m \}$
 gives all the remaining time instants at which the sensors  can be activated. We enumerate all the indices in the union to determine the index $j^*$ such that the objective function of \eqref{eq: prob_sch} is minimized as $w_{j^*} = 1$. 
We   summarize the greedy algorithm for non-myopic sensor scheduling in Algorithm\,4.}

\begin{algorithm}
\caption{Greedy algorithm for  sensor scheduling}
\begin{algorithmic}[1]
\Require $\mathbf w = \mathbf 0$ and $ {\mathcal I}_i = \{ i, i +m, \ldots, i +(\tau -1)m\}$ for $i = 1,2,\ldots, m$
\For {$l = 1, 2, \ldots, \min\{ s, \sum_{i=1}^m s_i \}$}
\State if $\tau - | \mathcal I_i | \geq s_i$, then replace $\mathcal I_i$ with an empty set for 
\hspace*{0.21in}$i = 1,2,\ldots, m$,
\State enumerate indices of $\mathbf w$ in $\{ {\mathcal I}_1 \cup {\mathcal I}_2 \cup \ldots \cup {\mathcal I}_m \}$ to 
\hspace*{0.21in}select $j^*$ such that   the  objective function of \eqref{eq: prob_sch}  is  
\hspace*{0.21in}minimized when $w_j = 1$,
\State remove $j$ from ${\mathcal I}_{i^*}$, where   $i^* $ is given by the remainder 
\hspace*{0.21in}of $\frac{j}{m}$   for  $i^* \neq  m $, and $i^* =  m $ if the remainder is $0$.
\EndFor
\end{algorithmic}
\end{algorithm}

The computational complexity of Algorithm\,4 is dominated by 
Step\,3. Specifically,   we evaluate the objective function of  \eqref{eq: prob_sch} using $O(\tau m)$ operations. And the computation of the Fisher information matrix requires a  
complexity of $O(\tau m^{2.373})$, where $O(\tau)$ accounts for the number of recursions,  and  $O( m^{2.373})$  is the complexity of matrix inversion in \eqref{eq: Gt_particle} \cite{Williams12}. 
We emphasize that different from Proposition\,\ref{prop: FIM}, expressing the closed form of the performance improvement in a greedy manner becomes intractable, since the Fisher information matrices are coupled with each other over the time horizon.
Therefore, 
the    computation cost of Algorithm\,4 is given by $O(\tau^2m^{3.373} )$ per iteration. 

For additional perspective, we compare the computational
complexity of Algorithm\,4 with the method in \cite{YRB2011},
where  a reweighted $\ell_1$ based quadratic programming (QP) was used to obtain locally optimal sensor schedules {under linear (or linearized) dynamical systems with correlated noise}. 
It was shown in \cite{YRB2011} that
the computational complexity of QP was ideally given by 
$O(m^{2.5}\tau^5)$ for every reweighting $\ell_1$ iteration. 
We note that  the computational complexity of the greedy algorithm   increases slightly  in terms of the network size by a factor $m^{0.873}$, while it
  decreases significantly  in terms of the length of the time horizon  by a factor $\tau^3$.

\section{Numerical Results}
\label{sec: num}
In this section, we demonstrate the effectiveness of the proposed approach  for sensor selection/scheduling with correlated measurement noise. In our numerical examples, we assume that   the sensors are randomly deployed in a  square region, where each of them provides the measurement of an  unknown parameter or state. 
For parameter estimation, we use the linear MMSE estimator  \cite[Sec.\,12]{karbook} to estimate the unknown parameter. For state tracking, we use the extended Kalman filter \cite[Sec.\,13]{karbook} to track the  target state. 

\subsubsection*{Sensor selection for parameter estimation}

We consider a network with $m \in \{ 20, 50\}$ sensors to estimate the vector of parameters $\mathbf x \in \mathbb R^n$ with $n = 2$, {where  sensors are randomly deployed  over a $50 \times 50$ lattice.}
The prior PDF of $\mathbf x$ is given by $\mathbf x \sim \mathcal N(\boldsymbol \mu, \boldsymbol \Sigma)$, where  $\boldsymbol \mu = [10, 10]^T$ and $\boldsymbol \Sigma = \mathbf I$. 
For simplicity, the row vectors of the measurement matrix $\mathbf H$  are chosen randomly, and independently, from the
distribution $\mathcal N(\mathbf 0, \mathbf I/\sqrt{n})$ \cite{SS2009}.
The covariance matrix of the measurement noise is set by an exponential model  \cite{liumasfarvar14_cf}{
\begin{align}
  R_{ij} = \cov(v_{i}, v_{j}) 
 = \sigma_v^2 \, 
 e^{- \rho  \| \boldsymbol \beta_i -\boldsymbol \beta_j \|_2 }, \label{eq: Psi_exp}
\end{align}
for $i,j = 1,2,\ldots,m$, where $\sigma_v^2 = 1$,   $\boldsymbol \beta_i \in \mathbb R^2$ is the location of the $i$th sensor in the 2D plane, $\| \cdot \|_2$ denotes the Euclidean norm,}
and $\rho $ is the correlation parameter which 
governs the strength of spatial correlation, namely, a larger (or smaller) $\rho$ corresponds to a weaker (or stronger) correlation.

We choose $N = 100$ while performing the   randomization method.
Also, we  employ  an exhaustive search that enumerates all    possible sensor selection schemes to obtain the globally optimal solution of  \eqref{eq: prob_sel}. The estimation performance is measured through the empirical MSE, which is averaged over $1000$ numerical trials.

\begin{figure}[htb]
\centerline{ \begin{tabular}{c}
\includegraphics[width=.5\textwidth,height=!]{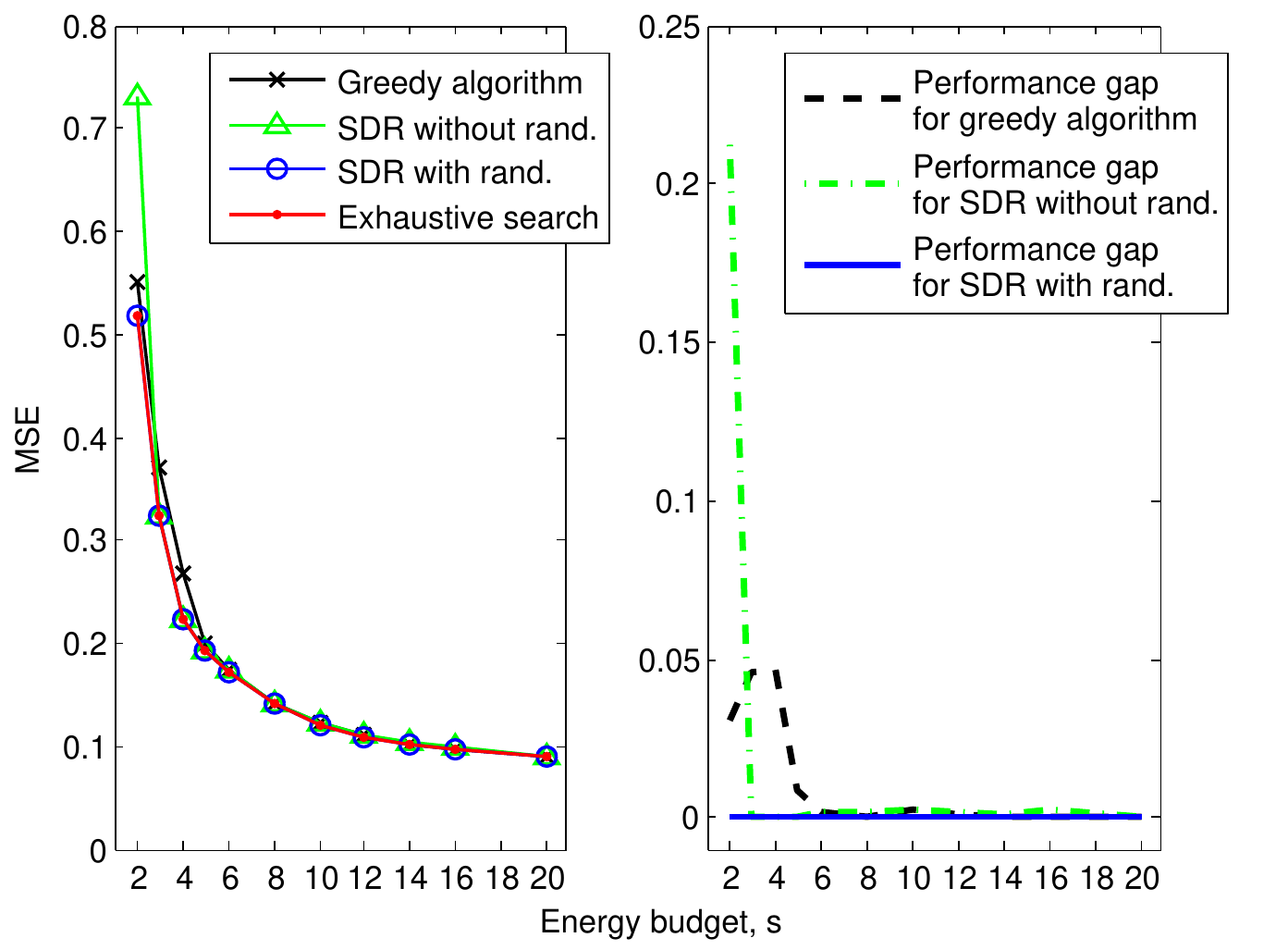}\\
(a) \\
\hspace*{-0.1in}\includegraphics[width=.5\textwidth,height=!]{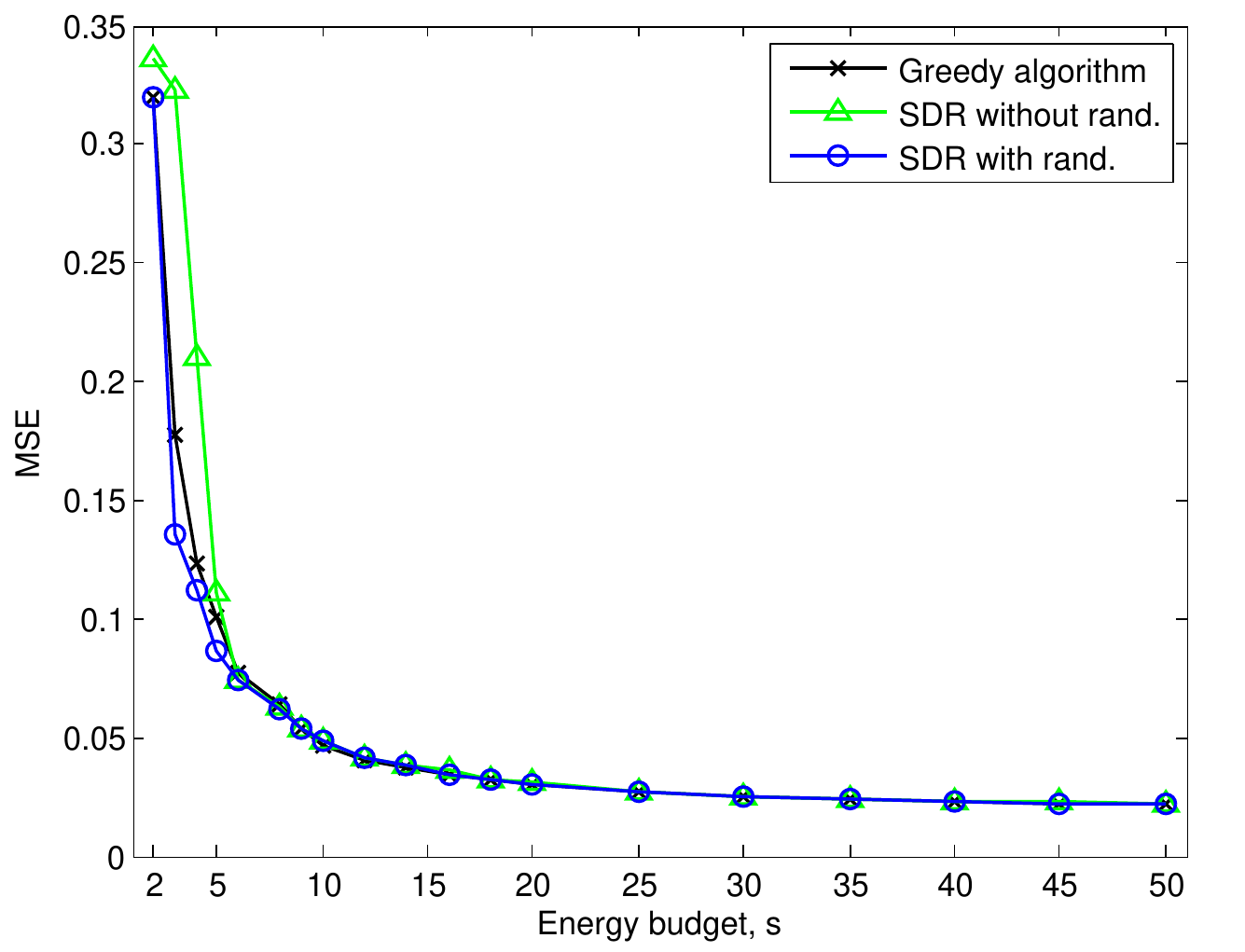}
\\
(b)
\end{tabular}}
\caption{\footnotesize{
MSE versus energy budget  with  correlation parameter $\rho = 0.1$.
}}
  \label{fig: MSE_ene_gen}
\end{figure}
 
In Fig.\,\ref{fig: MSE_ene_gen}, we present the MSE as  a function  of the energy budget  by solving \eqref{eq: prob_sel} with    correlation parameter $\rho = 0.1$. 
In Fig.\,\ref{fig: MSE_ene_gen}-(a) for the tractability of  exhaustive
search, we consider a small network with $m = 20$ sensors.
We compare the performance of the proposed greedy algorithm and SDR with randomization to that of 
SDR without randomization and exhaustive search. In particular, the right plots of Fig.\,\ref{fig: MSE_ene_gen}-(a) show the  performance gaps for the obtained locally optimal solutions compared to the  globally optimal solutions resulting from an exhaustive search. We observe that the SDR method with randomization outperforms the greedy algorithm and yields optimal solutions. 
The randomization method also significantly improves the performance of SDR in sensor selection. This is not surprising, and our numerical
observations agree with the literature \cite{luomasoyezha10,aspboy03report} that
demonstrate  the power and utility of randomization in SDR.

In Fig.\,\ref{fig: MSE_ene_gen}-(b), we present the MSE as a function of the energy budget for a relatively large network ($m = 50$). Similar to the results of Fig.\,\ref{fig: MSE_ene_gen}-(a), 
 the   SDR method with randomization  yields the lowest estimation error. 
We also observe that the MSE ceases to decrease significantly when $s \geq 20$. This indicates that  a subset of sensors suffices to provide satisfactory estimation performance, since the presence of correlation among sensors introduces  information redundancy and makes  observations less diverse.

In Fig.\,\ref{fig: MSE_ene_weak}, we solve the problem of sensor selection  
with weak noise correlation ($\rho = 0.5$), and present 
 the  MSE as a function of the energy budget  $s \in \{  2,3,\ldots,50\}$.  We compare the performance of three optimization approaches:
SDR with randomization for solving \eqref{eq: prob_weak}, bilinear programming (BP) for solving \eqref{eq: prob_weak_trJ}, and SDR with randomization for solving 
\eqref{eq: prob_weak_trJ}. We recall that  \eqref{eq: prob_weak} is to minimize the trace of the error covariance matrix and  \eqref{eq: prob_weak_trJ} is to maximize the trace of Fisher information.
As we can see, 
approaches that maximize the trace of  Fisher information   yield worse estimation performance than those that
minimize the estimation error. This is because \eqref{eq: prob_weak_trJ} ignores the contribution of prior information $\boldsymbol \Sigma$ in sensor selection.
We also note that although
BP (a linear programming based approach) has the lowest computational complexity, it leads to the worst optimization performance.

\begin{figure}[htb]
\includegraphics[width=.5\textwidth,height=!]{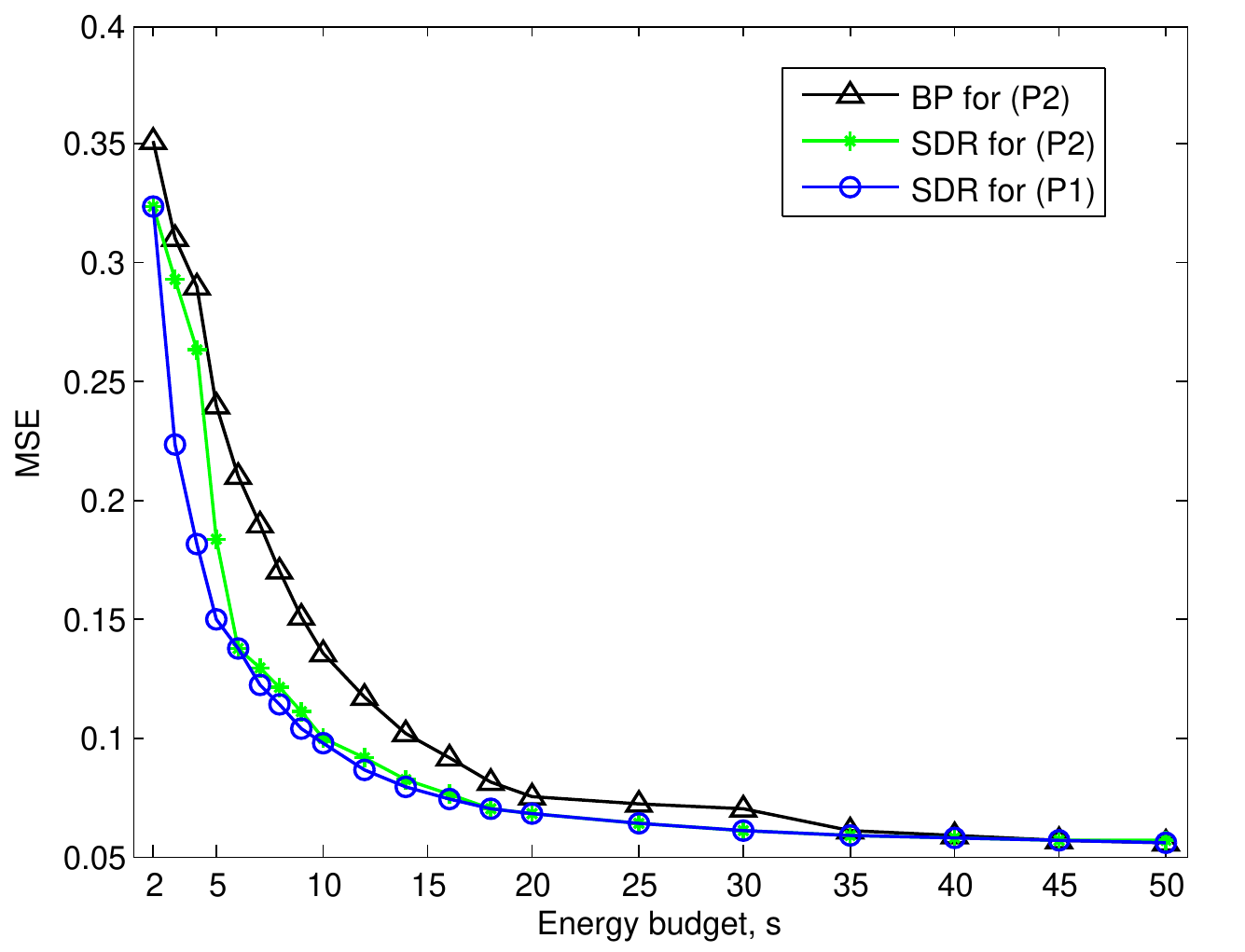} 
\caption{\footnotesize{MSE versus energy budget for sensor selection with weak noise correlation $\rho = 0.5$.
}}
  \label{fig: MSE_ene_weak}
\end{figure}

\begin{figure}[htb]
\includegraphics[width=.48\textwidth,height=!]{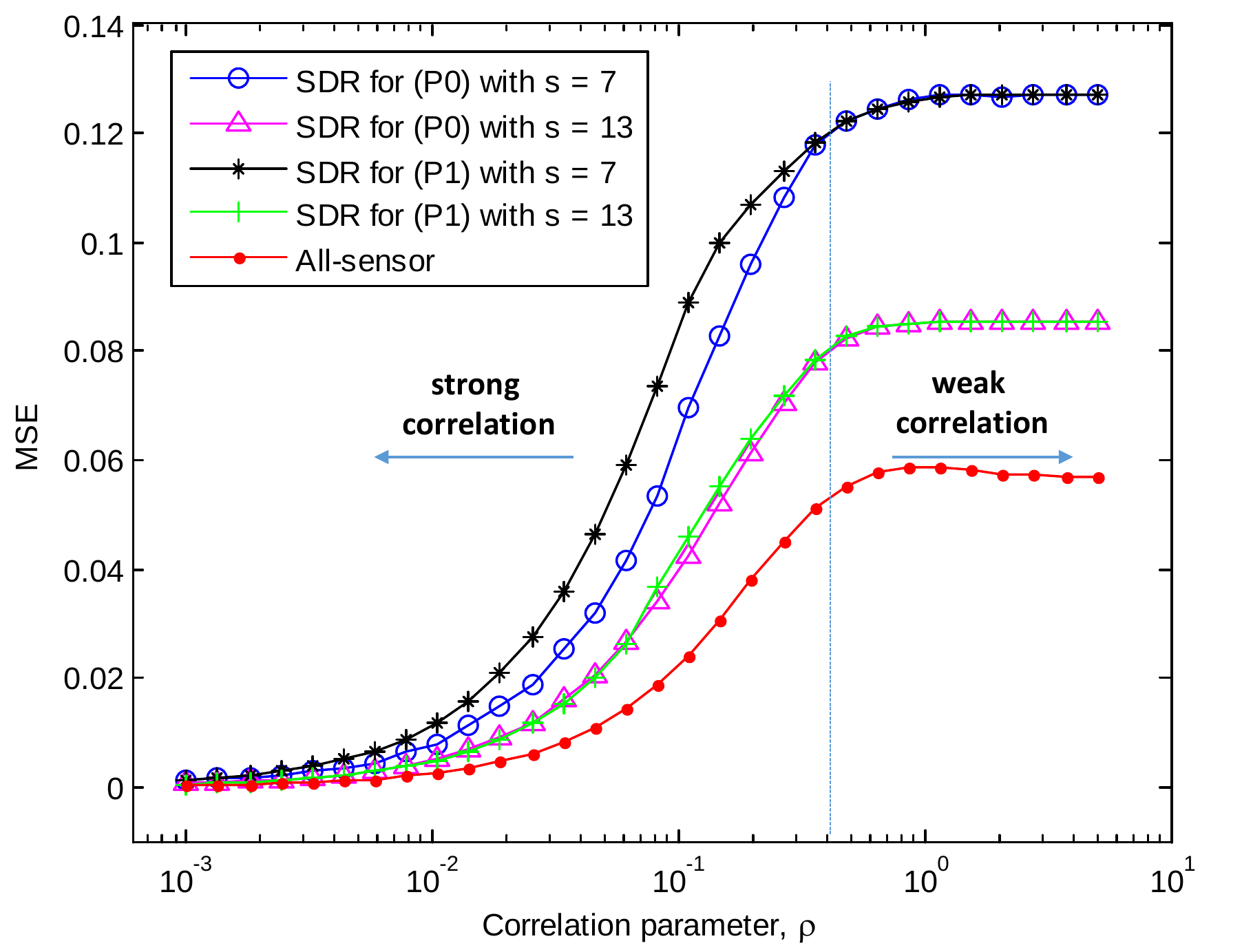} 
\caption{\footnotesize{MSE versus the strength of correlation for $s \in \{ 7, 13\}$.
}}
  \label{fig: MSE_corr}
\end{figure}

In Fig.\,\ref{fig: MSE_corr}, we present the MSE as a function of the correlation parameter $\rho$, where $m = 50$ and $s \in \{ 7, 13\}$. 
We consider   sensor selection schemes by using SDR with randomization to 
 solve problems \eqref{eq: prob_sel}  and
 \eqref{eq: prob_weak}, respectively.
For comparison, we also present the estimation performance when all the sensors are selected. 
As demonstrated in Fig.\,\ref{fig: MSE_corr}, we consider two  
  correlation regimes: weak correlation  and strong correlation. 
We  observe that in the weak correlation regime,    solutions of both
  \eqref{eq: prob_sel} and  \eqref{eq: prob_weak} yield   the same estimation performance.   
In   the strong correlation regime, 
solutions of \eqref{eq: prob_weak} could  lead to  worse estimation performance for sensor selection. 
We also observe that the sensitivity to the strategy of sensor selection reduces if the strength of correlation becomes extremely large, e.g., $\rho \leq 0.05$. More interestingly,
the estimation performance is improved as the correlation becomes stronger. 
This is because for strongly correlated noise, noise
cancellation  could be achieved by subtracting one observation from the other  \cite{penche13}. 
Further if we fix the value of $\rho$, the estimation error decreases when the energy budget increases, and the performance gap between solutions of  \eqref{eq: prob_sel} and  \eqref{eq: prob_weak} reduces.

\subsection*{Sensor scheduling for state tracking}

In this example, we track a target with $m = 30$ sensors over $30$ time steps.
We assume that the target state is 
a $4 \times 1$ vector $ \mathbf x_t = [x_{t,1}, x_{t,2}, x_{t,3}, x_{t,4}]^T$, 
where $(x_{t,1}, x_{t,2})$ and $(x_{t,3}, x_{t,4})$ denote the target location and velocity at time step $t$. 
The state equation \eqref{eq: state_gen} follows a white noise acceleration model \cite{masfarvar12}
\begin{equation*}
\mathbf F_t = \left[
  \begin{array}{cccc}
    1 & 0 & \Delta & 0 \\
    0 & 1 & 0 & \Delta \\
    0 & 0 & 1 & 0 \\
    0 & 0 & 0 & 1 \\
  \end{array}
\right], ~
 \mathbf Q = q \left[
  \begin{array}{cccc}
    \frac{\Delta^3}{3} & 0 & \frac{\Delta^2}{2} & 0 \\
    0 & \frac{\Delta^3}{3} & 0 & \frac{\Delta^2}{2} \\
    \frac{\Delta^2}{2} & 0 & \Delta & 0 \\
    0 & \frac{\Delta^2}{2} & 0 & \Delta \\
  \end{array}
\right],
\end{equation*}
where $\Delta$ and $q$ denote the sampling interval and the process noise
parameter, respectively. In our simulations, 
we set $\Delta = 1$ and $q = 0.01$. The prior PDF of the initial state is assumed to be Gaussian with mean $\hat {\mathbf x}_0 = [1,1,0.5,0.5]^T$ and covariance $\hat {\boldsymbol \Sigma}_0 = \mathrm{diag}(1, 1, 0.1,0.1)$. 
The measurement equation   follows
a   power attenuation model \cite{NV2006},
\begin{align}
h_{i} (\mathbf x_t) &= \sqrt{\frac{P_0}{1+ (x_{t,1} - \beta_{i,1})^2 + (x_{t,2} - \beta_{i,2})^2}}
\label{eq: h_meas}
\end{align}
for $i = 1,2,\ldots, m$, where  $P_0 = 10^4$ is the signal power of the source, and  the pair $(\beta_{i,1}, \beta_{i,2})$ is the position of the $i$th sensor. The covariance matrix of the measurement noise is given by \eqref{eq: Psi_exp} with $\rho = 0.035$.

\begin{figure}[htb]
\includegraphics[width=.47\textwidth,height=!]{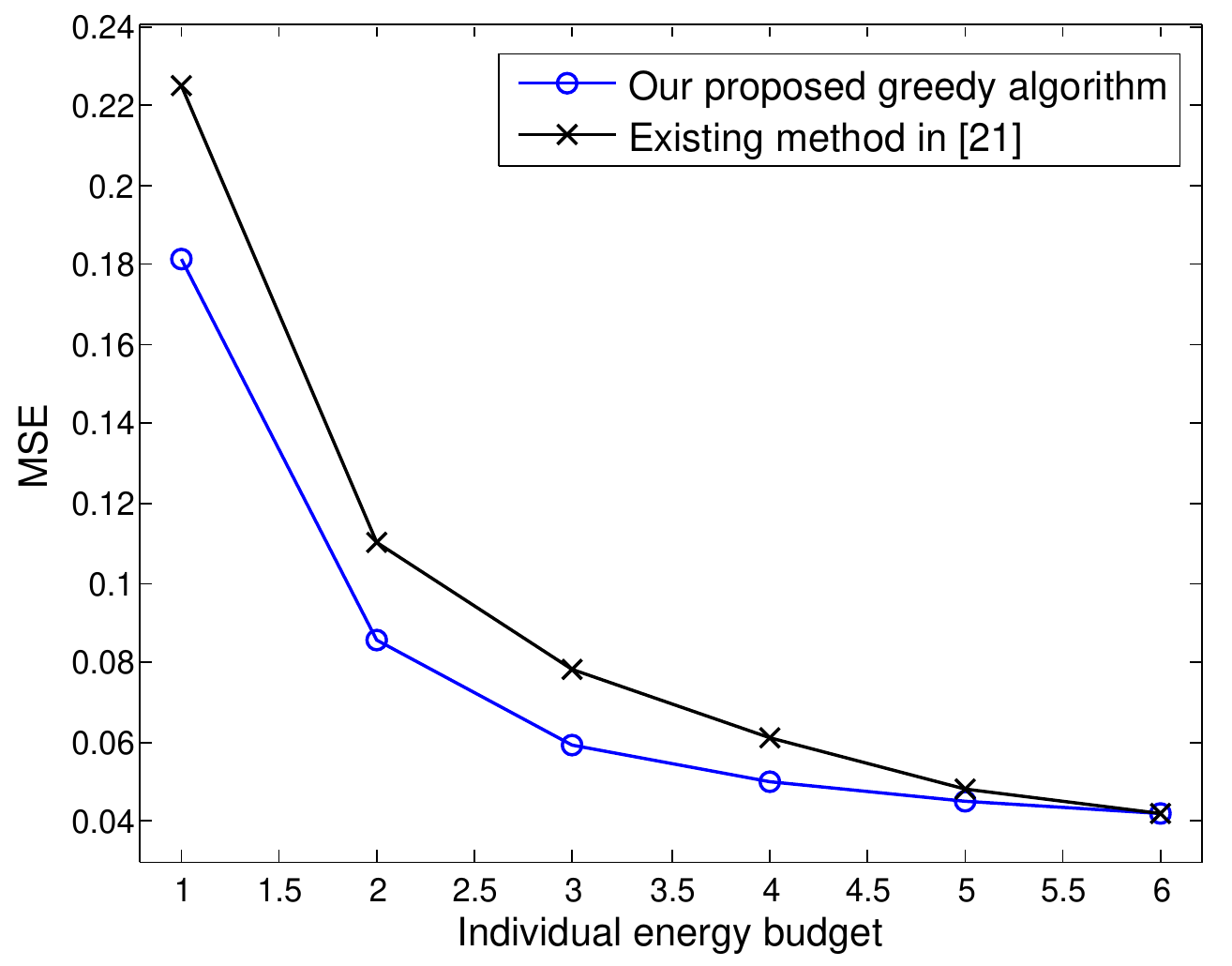} 
\caption{\footnotesize{MSE versus individual energy budget in target tracking.
}}
  \label{fig: MSE_ene_track}
\end{figure}

\begin{figure*}[htb]
\centerline{ \begin{tabular}{cc}
\includegraphics[width=.45\textwidth,height=!]{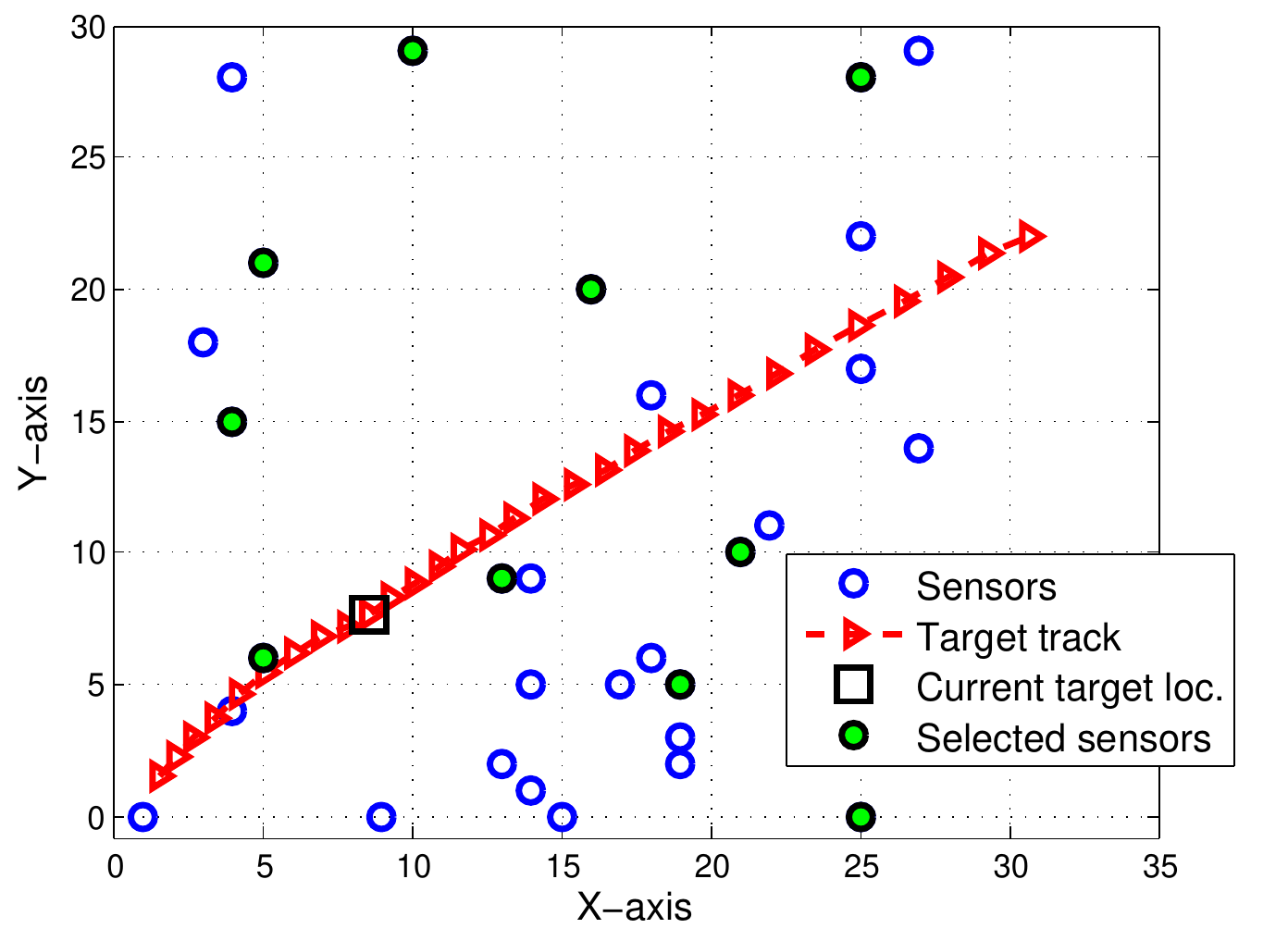}
& \includegraphics[width=.45\textwidth,height=!]{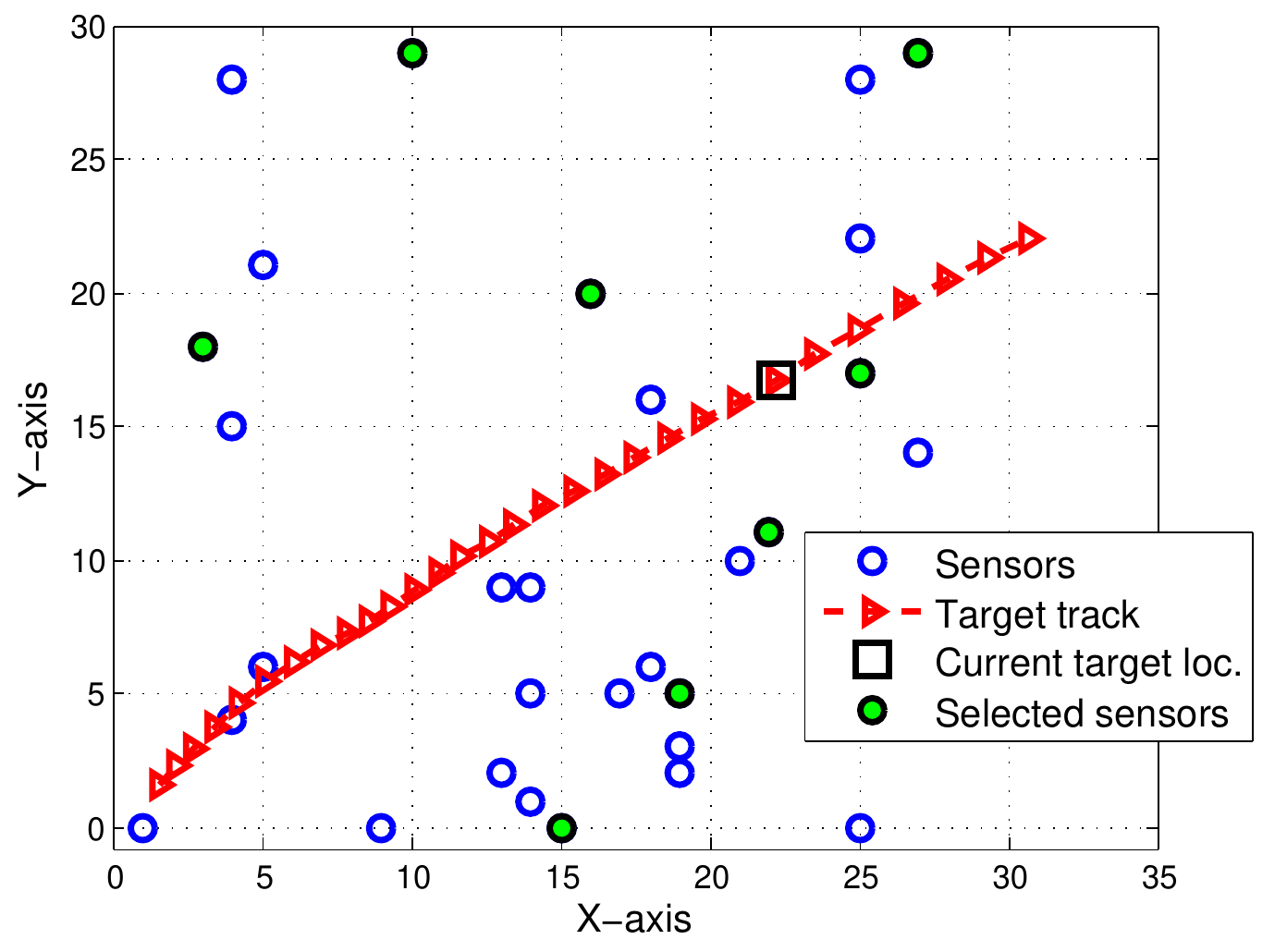}
\\
(a) & (b)
\end{tabular}}
\caption{\footnotesize{
Sensor schedules when $s_i = 2$: (a) $t = 10$, (b) $t = 24$.
}}
  \label{fig: schedule}
\end{figure*}

In  the sensor scheduling problem  \eqref{eq: prob_sch}, we assume  $s = \sum_{i = 1}^m s_i$ and $s_1 = s_2 = \cdots = s_m$.  
{In order to implement the proposed greedy algorithm and the existing   method in \cite{YRB2011},   the nonlinear measurement function \eqref{eq: h_meas}  is linearized   at the prediction state $\hat {\mathbf x}_t  = \mathbf F_{t-1} \mathbf F_{t-2}   \cdots   \mathbf F_0   \hat{\mathbf x}_0$ as   suggested in Remark\,\ref{Remark: nonlinear}.}
We  determine sensor schedules for every   $\tau = 6$ future time steps, and then update the estimate of the target state based on 
the {selected} measurements via an extended Kalman filter \cite{yushesah04}.
The estimation performance is measured through the empirical MSE, which  is obtained by averaging the estimation error  over $30$ time steps and $1000$ simulation trials.

In Fig.\,\ref{fig: MSE_ene_track}, we present the MSE as a function of the individual energy budget.
We compare the performance of our proposed greedy algorithm with that of 
 the sensor scheduling method in \cite{YRB2011}.
We remark that the method in \cite{YRB2011} relies on a reformulation of {linearized} dynamical systems and an $\ell_1$  relaxation  in optimization. In contrast, the proposed greedy algorithm  is independent of the dynamical system models and  convex relaxations. 
We   observe that
the greedy algorithm outperforms the method in \cite{YRB2011}. 
This result together with 
the previous results in Fig.\,\ref{fig: MSE_ene_gen} and \ref{fig: MSE_ene_weak} have implied that the greedy algorithm could yield  satisfactory estimation performance. 
Sensor schedules  at time steps $t = 10$ and $24$ are shown in Fig.\,\ref{fig: schedule}. 
We observe that some
 sensors closest to the target are selected due to their high  
 signal  power. 
However, from the entire network point of view, 
 the active sensors tend to be spatially
distributed rather than aggregating in a small neighborhood around
the target. This is because observations from neighboring sensors are strongly   correlated in space and may lead to information redundancy in target tracking.

\section{Conclusion}
\label{sec: conc}

In this paper, we studied the problem of sensor selection/scheduling with correlated measurement noise. 
We proposed
a  general  but tractable framework to design optimal sensor activations. 
We pointed out some drawbacks of   the existing frameworks for sensor selection with correlated noise, and showed that the existing formulation is  valid only  for the special case of weak noise correlation.
Further, we extended our framework to the problem of non-myopic sensor scheduling, where a greedy algorithm was developed to design non-myopic sensor schedules. Numerical results were  provided to illustrate the effectiveness of our  approach  and the impact of noise correlation on the performance of sensor selection.

In future work, we will  study  applications of sensor selection with correlated noise, such as localization in multipath environments,  sensor collaboration in distributed estimation, and clock synchronization in wireless sensor networks.
{It would also be of interest to seek theoretical guarantees for the   performance of the greedy algorithm.}
Furthermore,
in order to reduce
the computational burden at the fusion center, developing
a decentralized architecture where the optimization procedure
can be carried out in a distributed way and by the sensors
themselves is another direction of future research.

\appendices

\section{Proof of Proposition\,\ref{prop: FIM}}
\label{app:prop1}
{
Given the sensor selection scheme $\tilde{\mathbf w}$, it is clear from \eqref{eq: Jw} that Fisher information can be written as
\begin{align}
\hspace*{-0.03in} \mathbf J_{\tilde w}  = \boldsymbol \Sigma^{-1} \hspace*{-0.03in}+ [\mathbf H_w^T , \mathbf h_j] \mathbf R_v^{-1} \begin{bmatrix}
\mathbf H_w \\
\mathbf h_j^T
\end{bmatrix}, ~
\mathbf R_{\tilde w} \Def \begin{bmatrix}
\mathbf R_w & \mathbf r_j \\
\mathbf r_j^T & r_{jj}
\end{bmatrix}
\label{eq: Jv}
\end{align}
where {$\mathbf H_w \Def \boldsymbol \Phi_w \mathbf H$}.

If $\mathbf w \neq \mathbf 0$, the  inverse of $\mathbf R_{\tilde w}$ in 
\eqref{eq: Jv} is given by
\begin{align}
\mathbf R_{\tilde w}^{-1} = c_j
\begin{bmatrix}
{c_j}^{-1} \mathbf R_w^{-1} + \mathbf R_w^{-1} \mathbf r_j \mathbf r_j^T \mathbf R_w^{-1}  & - \mathbf R_w^{-1} \mathbf r_j \\
- \mathbf r_j^T \mathbf R_w^{-1} & 1
\end{bmatrix}
\label{eq: Rv_inv}
\end{align}
where $c_j \Def 1/(r_{jj} - \mathbf r_j^T \mathbf R_w^{-1} \mathbf r_j)
$, and $c_j >0$ following from the Schur complement of $\mathbf R_{\tilde w}$.
Substituting \eqref{eq: Rv_inv} into \eqref{eq: Jv}, we obtain
\begin{align}
\mathbf J_{\tilde w} = \mathbf J_w + c_j \boldsymbol \alpha_j \boldsymbol \alpha_j^T,
\label{eq: Jvw}
\end{align}
where  $\mathbf J_w = \boldsymbol \Sigma^{-1} + \mathbf H_w^T \mathbf R_{w}^{-1} \mathbf H_w$ as indicated by \eqref{eq: Jw}, and $\alpha_j \Def \mathbf H_w^T \mathbf R_w^{-1} \mathbf r_j - \mathbf h_j$. 

If $\mathbf w = \mathbf 0$, namely, $\mathbf J_w = \boldsymbol \Sigma^{-1}$, we can immediately obtain from \eqref{eq: Jv} that 
\begin{align}
\mathbf J_{\tilde w} = \mathbf J_w + \frac{1}{r_{jj}} \mathbf h_j^T \mathbf h_j.
\label{eq: Jvw0}
\end{align}
 Equations  \eqref{eq: Jvw} and \eqref{eq: Jvw0} imply that $\mathbf J_{\tilde w} - \mathbf J_w \succeq \mathbf 0$ since $c_j > 0$.
 
 We apply the matrix inversion lemma to \eqref{eq: Jvw}. This yields
\begin{align*}
\mathbf J_{\tilde w}^{-1} & = [\mathbf J_w + c_j \boldsymbol \alpha_j \boldsymbol \alpha_j^T]^{-1}  = \mathbf J_w^{-1} - \frac{c_j \mathbf J_w^{-1} \boldsymbol \alpha_j \boldsymbol \alpha_j^T \mathbf J_w^{-1} }{1 + c_j \boldsymbol \alpha_j \mathbf J_w^{-1} \boldsymbol \alpha_j}.
\end{align*}
The improvement in estimation error is then given by
\begin{align*}
\tr(\mathbf J_w^{-1}) - \tr(\mathbf J_{\tilde w}^{-1})
= \frac{c_j \boldsymbol \alpha_j^T \mathbf J_w^{-2} \boldsymbol \alpha_j}{1 +c_j \boldsymbol \alpha_j \mathbf J_w^{-1} \boldsymbol \alpha_j }. 
\end{align*}
 \hfill $\blacksquare$}

 \section{Proof of Proposition\,\ref{prop: weak}}
\label{prob: Jweak}
Our goal is to simplify the  Fisher information matrix given by \eqref{eq: Jw} 
under the assumption of weak noise correlation.
According to \eqref{eq: R_weak}, we obtain
\begin{align}
\mathbf R_w^{-1} = &( \boldsymbol \Phi_{w}  \mathbf R \boldsymbol \Phi_w^T)^{-1} \nonumber \\
 = &  ( \boldsymbol \Phi_{w}  \boldsymbol \Lambda \boldsymbol \Phi_w^T + \epsilon  \boldsymbol \Phi_{w}  \boldsymbol \Upsilon \boldsymbol \Phi_w^T  )^{-1} \nonumber \\
\overset{(1)}{=}  & 
(\mathbf I + \epsilon \boldsymbol \Phi_{w}  \boldsymbol \Lambda^{-1}\boldsymbol \Phi_w^T  \boldsymbol \Phi_{w}  \boldsymbol \Upsilon \boldsymbol \Phi_w^T )^{-1} \boldsymbol \Phi_{w}  \boldsymbol \Lambda^{-1}\boldsymbol \Phi_w^T \nonumber \\
\overset{(2)}{=}  &
(\mathbf I - \epsilon \boldsymbol \Phi_{w}  \boldsymbol \Lambda^{-1}\boldsymbol \Phi_w^T  \boldsymbol \Phi_{w}  \boldsymbol \Upsilon \boldsymbol \Phi_w^T) \boldsymbol \Phi_{w}  \boldsymbol \Lambda^{-1}\boldsymbol \Phi_w^T \nonumber \\
&+ O(\epsilon^2)  \quad  (\text{as $\epsilon \to 0$})\nonumber \\
\overset{(3)}{=} & \boldsymbol \Phi_{w}  \boldsymbol \Lambda^{-1}\boldsymbol \Phi_w^T
- \epsilon \boldsymbol \Phi_{w}  \boldsymbol \Lambda^{-1} \mathbf D_w  \boldsymbol \Upsilon \mathbf D_w \boldsymbol \Lambda^{-1}\boldsymbol \Phi_w^T \nonumber \\
& + O(\epsilon^2)  \quad  (\text{as $\epsilon \to 0$}),
\label{eq: RD}
\end{align}
where $\mathbf D_w \Def \diag(\mathbf w)$.
In \eqref{eq: RD},  step (1) holds since we use the facts that $\boldsymbol \Lambda$ is a diagonal matrix and $(\boldsymbol \Phi_{w}  \boldsymbol \Lambda \boldsymbol \Phi_w^T)^{-1} = \boldsymbol \Phi_{w}  \boldsymbol \Lambda^{-1}\boldsymbol \Phi_w^T$; step (2) is obtained from the Taylor series expansion  $(\mathbf I + \epsilon \mathbf X)^{-1}  = \sum_{i=0}^{\infty} (-\epsilon \mathbf X)^{i}$  as $\epsilon \to 0$ (namely, the spectrum of $\epsilon \mathbf X$ is contained inside the open unit disk); step (3) is true since $\boldsymbol \Phi_{w}^T \boldsymbol \Phi_{w} = \mathbf D_w$   as in \eqref{eq: Phi_property}.

Substituting \eqref{eq: RD} into \eqref{eq: Jw}, we obtain
\begin{align*}
\mathbf J_w = & \boldsymbol \Sigma^{-1} + \mathbf H^T \boldsymbol \Phi_w^T \boldsymbol \Phi_{w}  \boldsymbol \Lambda^{-1}\boldsymbol \Phi_w^T 
\boldsymbol \Phi_w \mathbf H \nonumber \\
& - \epsilon \mathbf H^T \boldsymbol \Phi_w^T  \boldsymbol \Phi_{w}  \boldsymbol \Lambda^{-1} \mathbf D_w  \boldsymbol \Upsilon \mathbf D_w \boldsymbol \Lambda^{-1}\boldsymbol \Phi_w^T \boldsymbol \Phi_w \mathbf H \nonumber\\ 
&+  O(\epsilon^2)   \quad  (\text{as $\epsilon \to 0$}) \nonumber \\
\overset{(1)}{=}  &  \boldsymbol \Sigma^{-1} + \mathbf H^T  (\mathbf D_w \boldsymbol \Lambda^{-1} \mathbf D_w - \epsilon \mathbf D_w \boldsymbol \Lambda^{-1} \boldsymbol \Upsilon \boldsymbol \Lambda^{-1} \mathbf D_w)  \mathbf H \nonumber \\
& + O(\epsilon^2)  \quad  (\text{as $\epsilon \to 0$}) \nonumber \\
=& \boldsymbol \Sigma^{-1} + \mathbf H^T \mathbf D_w  ( \boldsymbol \Lambda^{-1}   - \epsilon   \boldsymbol \Lambda^{-1} \boldsymbol \Upsilon \boldsymbol \Lambda^{-1}  ) \mathbf D_w \mathbf H \nonumber \\
& + O(\epsilon^2)   \quad  (\text{as $\epsilon \to 0$}) \nonumber \\
\overset{(2)}{=}  &    \boldsymbol \Sigma^{-1} + \mathbf H^T \mathbf D_w  \mathbf R^{-1} \mathbf D_w \mathbf H + O (\epsilon^2)   \quad  (\text{as $\epsilon \to 0$}) \nonumber \\
\overset{(3)}{=} & \boldsymbol \Sigma^{-1} + \mathbf H^T(\mathbf w \mathbf w^T \circ \mathbf R^{-1}) \mathbf H + O (\epsilon^2)  \quad  (\text{as $\epsilon \to 0$}), 
\end{align*}
where step (1) is achieved by using the fact that 
$\mathbf D_w \boldsymbol \Lambda^{-1} = \boldsymbol \Lambda^{-1} \mathbf D_w = \mathbf D_w \boldsymbol \Lambda^{-1} \mathbf D_w$,   step (2) holds due to 
$\mathbf R^{-1}  =  \boldsymbol \Lambda^{-1}   - \epsilon  \boldsymbol \Lambda^{-1} \boldsymbol \Upsilon \boldsymbol \Lambda^{-1}  + O(\epsilon^2) $, and step (3) is true since 
$\mathbf D_w$ is diagonal and has only
binary elements.
 \hfill $\blacksquare$

\section{Proof of Proposition\,\ref{prop: trJ}}
\label{app: prop_trJ}
We begin by simplifying the objective function in \eqref{eq: prob_weak_trJ},
\begin{align}
\phi(\mathbf w) \Def & \tr(\boldsymbol \Sigma^{-1}) +  \tr \left ( (\mathbf w \mathbf w^T \circ \mathbf R^{-1}) (\mathbf H^T \mathbf H)  \right ) \nonumber \\
=&\tr(\boldsymbol \Sigma^{-1}) +
\sum_{i=1}^m \sum_{j=1}^m w_i w_j \bar {R}_{ij} \mathbf h_i^T \mathbf h_j \nonumber \\
= & \tr(\boldsymbol \Sigma^{-1}) + \mathbf w^T \boldsymbol \Omega \mathbf w, \label{eq: obj_compact}
\end{align}
where $\bar  { R}_{ij}$ is the $(i,j)$th entry of $ {\mathbf R}^{-1}$,    and $
\bar {R}_{ij} \mathbf h_i^T \mathbf h_j 
$ corresponds to the $(i,j)$th entry of $\boldsymbol \Omega$ which yields the  succinct form
\begin{align}
\boldsymbol \Omega = \mathbf A (\mathbf R^{-1} \otimes \mathbf I_{n }) \mathbf A^T. \label{eq: Omega_compact}
\end{align}
In \eqref{eq: Omega_compact},
$\otimes$
denotes the Kronecker product,  
$
\mathbf A \in \mathbb R^{m \times mn}
$ is a block-diagonal matrix whose diagonal blocks are given by $\{ \mathbf h_i^T \}_{i=1}^m$, 
and $\boldsymbol \Omega \succeq 0$ due to
$
\mathbf R^{-1} \otimes \mathbf I_{n } \succeq 0
$.

According to  \eqref{eq: obj_compact}, \eqref{eq: prob_weak_trJ} can be rewritten as
\begin{align}
\begin{array}{ll}
\displaystyle \maximize_{\mathbf w} & ~  \mathbf w^T \boldsymbol \Omega \mathbf w\\
\st &~ \mathbf 1^T \mathbf w \leq s, \\
&~  \mathbf w \in \{ 0,1\}^m.
\end{array}
\label{eq: prob_weak_w_quad}
\end{align}

Next, we prove that problem \eqref{eq: prob_weak_w_relax} is equivalent to problem \eqref{eq: prob_weak_w_quad}. We recall that
the former is a relaxation of the latter, where the former entails the maximization of a convex quadratic function over a bounded polyhedron
$
\mathcal P \Def \{ \mathbf w | \mathbf 1^T \mathbf w \leq s,  \mathbf w \in [0,1]^m\}
$.
It has been shown in \cite{kon76} that optimal solutions of such a problem   occur at vertices of the polyhedron $\mathcal P$, which are zero-one vectors. This indicates that    solutions of  problem \eqref{eq: prob_weak_w_relax}  are  feasible for problem \eqref{eq: prob_weak_w_quad}.  
{
Therefore, solutions of  \eqref{eq: prob_weak_w_relax} are
solutions of  \eqref{eq: prob_weak_w_quad}, and vice versa.}
\hfill $\blacksquare$

\ifCLASSOPTIONcaptionsoff
  \newpage
\fi



%

\bibliographystyle{IEEEbib}
\bibliography{journal,journal_col}

\end{document}